\documentclass[aps,prd,onecolumn,preprintnumbers,notitlepage,superscriptaddress,
nofootinbib,amsfont,amssymb,10pt,singlespacing] {revtex4-1}
\usepackage{amsmath,bm}
\usepackage{times}
\usepackage{braket}
\usepackage{color,graphicx}
\usepackage{slashed}
\usepackage{hyperref}
\usepackage{graphics}
\usepackage{subfigure}
\usepackage{multirow,makecell}
\usepackage{textcomp}
\usepackage{mathtools}
\usepackage{float}

\newcommand{\beq}{\begin{eqnarray}}
\newcommand{\eeq}{\end{eqnarray}}

\def\lsim{ {\ \lower-1.2pt\vbox{\hbox{\rlap{$<$}\lower6pt\vbox{\hbox{$\sim$}
}}}\ } }
\def\gsim{ {\ \lower-1.2pt\vbox{\hbox{\rlap{$>$}\lower6pt\vbox{\hbox{$\sim$}
}}}\ } }
\definecolor{Red}{rgb}{1.,0.,0.}

\definecolor{Blue}{rgb}{0.,0.,1.}

\definecolor{nicered}{rgb}{0.7,0.1,0.1}
\definecolor{nicegreen}{rgb}{0.1,0.5,0.1}

\hypersetup{colorlinks,citecolor=nicegreen,linkcolor=nicered}

\begin{document}

\title{
Investigating $B_{s}$ three-body decays of scalar mesons in perturbative QCD approach}
%%%==================================================================

\author{Hong~Yang}
\email[Electronic address:]{15116233293@163.com}
\affiliation{School of Physical Science and Technology,
 Southwest University, Chongqing 400715, China}

\author{Xian-Qiao~Yu}
\email[Electronic address:]{yuxq@swu.edu.cn}
\affiliation{School of Physical Science and Technology,
Southwest University, Chongqing 400715, China}

\date{\today}

%%%%%%%%%%%%%%%%%%%%%%%%%%%%%%%%%%%%%%%%%%%%%%%%%%%%%%%%%%%%%%%%%%
\begin{abstract}

In this study, we investigate the branching ratios of $B^{0}_{s}\rightarrow a_{0}(980)[ \rightarrow K\overline{K}, \pi\eta]a_{0}(980)$, $B^{0}_{s} \rightarrow f_{0}(980)[ \rightarrow \pi^{+}\pi^{-}, K^{+}K^{-}]f_{0}(980)$, and $B^{0}_{s} \rightarrow f_{0}(500)[ \rightarrow \pi^{+}\pi^{-}]f_{0}(500)$ decays in the pQCD approach, wherein the scalar mesons $a_{0}(980)$, $f_{0}(980)$, and $f_{0}(500)$ are regarded as the lowest-lying $q\overline{q}$ state. In the $SU(3)$ nonet, there exists a mixing between the scalars $f_{0}(980)$ and $f_{0}(500)$. Thus, we have considered the mixing effect in our calculations to obtain reliable data, and set the mixing angle $\theta$ as $[15^{\circ}, 82^{\circ}]$ and $[105^{\circ}, 171^{\circ}]$. Based on the isospin symmetry, we estimated the branching ratios of the $B^{0}_{s} \rightarrow f_{0}(980)[ \rightarrow \pi^{0}\pi^{0}]f_{0}(980)$ and $B^{0}_{s} \rightarrow f_{0}(500)[ \rightarrow \pi^{0}\pi^{0}]f_{0}(500)$ decays. The branching ratios of the $B^{0}_{s}\rightarrow a_{0}(980)[ \rightarrow K\overline{K}, \pi\eta]a_{0}(980)$ decays are much small, while those of the $B^{0}_{s} \rightarrow f_{0}(980)[ \rightarrow \pi\pi, K\overline{K}]f_{0}(980)$ and $B^{0}_{s} \rightarrow f_{0}(500)[ \rightarrow \pi\pi]f_{0}(500)$ decays are at the order of $10^{-6}\sim10^{-5}$, which can be tested in the LHCb and Belle II experiments, hopefully.

\end{abstract}
%%%%%%%%%%%%%%%%%%%%%%%%%%%%%%%%%%%%%%%%%%%%%%%%%%%%%%%%%%%%%%

\maketitle

%
%%%
%%%%%%%%%%%%%%%%% I. INTRODUCTION %%%%%%%%%%%%%%%%%%%%%%%%%%%%%%%%
%%%
%

\section{Introduction}\label{sec:intro}

The scalar mesons $a_{0}(980)$ and $f_{0}(980)$ have attracted immense attention since their discovery. These scalar mesons, as a key problem in the nonperturbative QCD~\cite{Achasov:2018ela}, play a crucial role in understanding the chiral symmetry and confinement in the low-energy region. However, the mysterious internal structure of the scalar mesons $a_{0}(980)$ and $f_{0}(980)$ remains a puzzle, many related researches have been carried out accordingly. It has been raised that the scalar mesons $f_{0}(500)$, $K^{*}(700)$, $f_{0}(980)$, and $a_{0}(980)$ form a $SU(3)$ flavor nonet, whereas the scalars above $1$ GeV, including $f_{0}(1370)$, $a_{0}(1450)$, $K^{*}(1430)$, and $f_{0}(1500)$, form a different nonet~\cite{P:2020mia}. In Ref.~\cite{Cheng:2006yov}, two scenarios have been proposed to describe the quark components of the light scalar mesons. According to the first scenario, the light scalar mesons contained in the first $SU(3)$ flavor nonet are the lowest-lying $q\overline{q}$ state. In the other scenario, the scalars in the second nonet are treated as the $q\overline{q}$ state, and the scalar mesons below or close to $1$ GeV are considered to be the four-quark bound state. In addition, the presence of nonstrange and strange quark contents in $f_{0}(980)$ and $f_{0}(500)$ have been confirmed experimentally; thus, they can be regarded as a mixture of $s\overline{s}$ and $(u\overline{u}+d\overline{d})/\sqrt{2}$~\cite{E:2001th}. The aforementioned studies provide a positive significance to explore the internal structure of the scalar mesons.

The perturbative QCD (pQCD) approach based on the $k_{T}$ factorization has been extensively used to study the decay of $B$ mesons~\cite{Yan:2018si,Zhou:2021si,Li:2022hao,Liu:2022hao,Wang:2014xyh}. It is well known that the QCD dynamics of the three-body decay is more complex than those of the two-body decay. In the pQCD approach, the three-body decay is usually simplified to a two-body decay by introducing two-hadron distribution amplitudes~\cite{M:1998si,M:2000xyh}, which can be called quasi-two-body decay, containing both resonance and nonresonance information. The dominant contributions are from the parallel motion region, where the invariant mass of the light meson pair is below $O(\overline{\Lambda}M_{B})$, and $\overline{\Lambda}=M_{B}-m_{b}$ is the difference in mass between $B$ meson and $b$-quark. Hence, the pQCD factorization formula for the three-body decay amplitude of the $B$ meson is written as~\cite{C:2004cpa,C:2003vda}

\begin{equation}
{\cal A}={\cal H}\otimes \phi_{B} \otimes \phi_{h_{3}} \otimes \phi_{h_{1}h_{2}},
\end{equation}
where the hard decay kernel ${\cal H}$ can be calculated by using the perturbative theory. The nonperturbative inputs $\phi_{B}$, $\phi_{h_{1}h_{2}}$ and $\phi_{h_{3}}$ are the distribution amplitudes of $B$ meson, $h_1h_2$ pair and $h_3$, respectively.

In the past few decades, several $B$ decays with a final state $a_{0}(980)$ or $f_{0}(980)$ have been observed in experiments~\cite{BABAR:2004ae,BABAR:2006siy,Belle:2006emv}, and the corresponding theoretical calculations have also attracted increased attention. In Ref.~\cite{C:2010wwdd,C:2011wwdd}, the $B_{s} \rightarrow f_{0}(980)$ transition form factors have been estimated, and the authors still predict the branching ratio of an interesting decay mode $B_{s} \rightarrow J/\psi f_{0}(980)$. At the same time, the light scalar mesons also can make contributions to the $B$ meson decays as intermediate resonance. For example, the LHCb Collaboration reported the $B^{0} \rightarrow J/{\psi}K^{+}K^{-}$ decay with the $a_{0}(980)$ resonance~\cite{Aaij:2013zpx}, and the Belle Collaboration observed the $B^{\pm} \rightarrow K^{\pm}f_{0}(980) \rightarrow K^{\pm}{\pi}^{\mp}{\pi}^{\pm}$ where the scalar meson $f_{0}(980)$ was regarded as the intermediate resonance~\cite{Belle:2002dka}. Recently, many works have been carried out to calculate the three-body decays of the $B$ meson with $a_{0}(980)$ or $f_{0}(980)$ resonance in the pQCD approach. In Ref.~\cite{R:2019ngy}, the branching ratios of the $B \rightarrow J/{\psi}(K\overline{K},\pi\eta)$ decays have been calculated with the contributions of the scalars $a_{0}(980)$ and $a_{0}(1450)$, in which the timelike form factors of $a_{0}(980)$ resonance and $a_{0}(1450)$ resonance are shaped by the Flatt$\acute{e}$ model and the Breit-Wigner formula, respectively. The authors of Ref.~\cite{J:2022wa}, employing $a_{0}(980)$, $a_{0}(1450)$ and $a_{0}(1950)$ as resonances, analyze the quasi-two-body $B \rightarrow a_{0}(\rightarrow K\overline{K},\pi\eta)h$ decays within the two scenarios mentioned in the first paragraph. In Ref.~\cite{W:2022uun} and Ref.~\cite{L:2021vb}, the branching ratios of the $B \rightarrow K({\cal R} \rightarrow K^{+}K^{-})$ and $B_{(s)} \rightarrow V\pi\pi$ decays with $f_{0}(980)$ resonance have been studied severally by considering the mixing of $s\overline{s}$ and $(u\overline{u}+d\overline{d})/\sqrt{2}$. Furthermore, the authors of Ref.~\cite{Z:2020aaa} explore the branching ratios and $CP$ violations of the two-body decay $B^{0}_{s} \rightarrow SS(a_{0}(980),f_{0}(980),f_{0}(500))$ in the pQCD approach for the first time, and the branching ratios are at the order of 10$^{-4}$ $\sim$ 10$^{-6}$ with a high probability to be tested experimentally in the future. Therefore, we further develop the work in Ref.~\cite{Z:2020aaa}. In our calculation, we regard the scalar mesons $a_{0}(980)$ and $f_{0}(980)$ as the $q\overline{q}$ state, just as mentioned in the first scenario, and predict the branching ratios of the quasi-two-body decays $B^{0}_{s} \rightarrow S[\rightarrow P_{1}P_{2}]S$, where $S$ denotes the light scalar mesons $a_{0}(980)$, $f_{0}(980)$ and $f_{0}(500)$ \footnote{$a_{0}$, $f_{0}$ and $\sigma$ refers to $a_{0}(980)$, $f_{0}(980)$ and $f_{0}(500)$ respectively in the following text}, and $P_{1}P_{2}=\pi\eta, \pi\pi, K\overline{K}$ is the final state meson pair. For the scalar mesons $f_{0}(980)$ and $f_{0}(500)$, we also adopt the corresponding mixing mechanism. The results presented in this paper can be validated in the LHCb and Belle II experiments in the near future.

The rest of this paper is organized as follows. In Section \ref{sec:pert}, the theoretical framework of the pQCD, the wave functions involved in the calculations and the helicity amplitudes for the $B^{0}_{s} \rightarrow S[\rightarrow P_{1}P_{2}]S$ decays are described. In Section \ref{sec:numer}, the numerical results are presented and discussed. In Section \ref{sec:summary}, a summary of this work is provided. And the explicit formulas of all the helicity amplitudes are presented in the Appendix.

\begin{figure}[htbp]
	\centering
	\begin{tabular}{l}
		\includegraphics[width=0.8\textwidth]{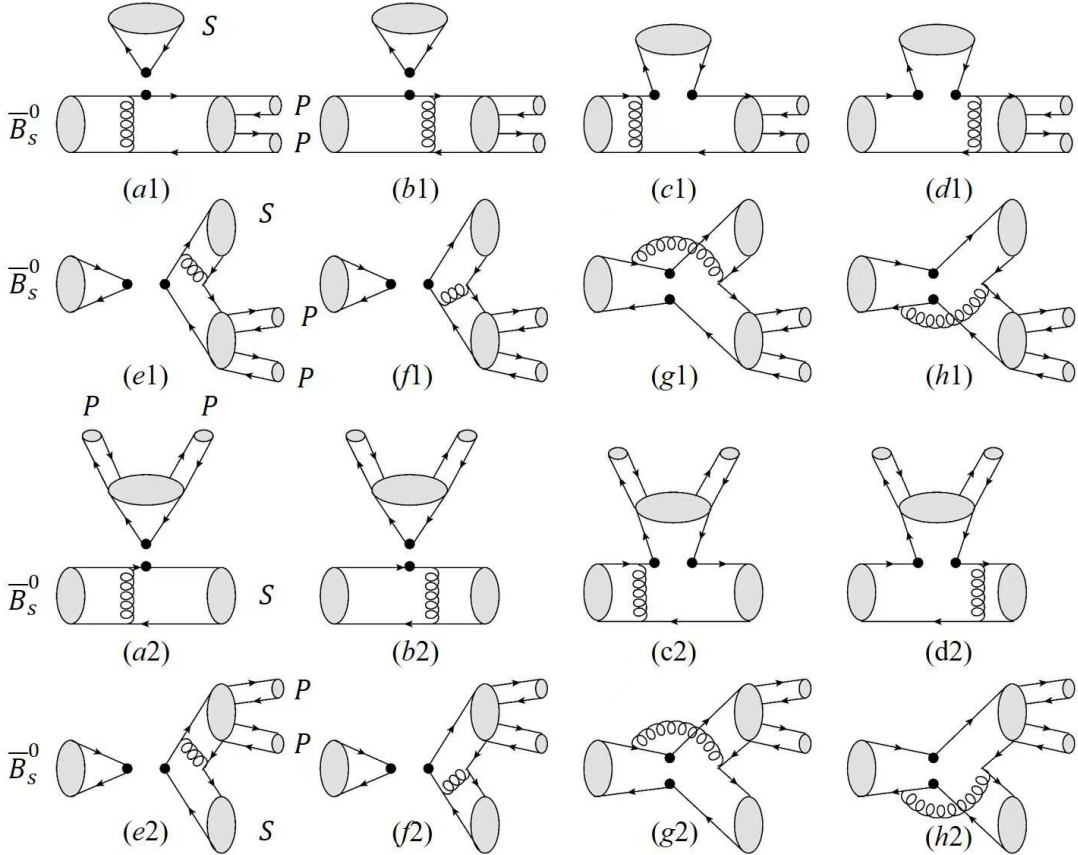}
	\end{tabular}
	\caption{The Feynman diagrams for the $\overline{B}^{0}_{s} \rightarrow S[S\rightarrow]{P_{1}P_{2}}$ decays in pQCD. The symbol $\bullet$ stands for the weak vertex, $S$ means the scalar mesons, and $P_{1}P_{2}$ denotes the final state meson pair, the corresponding relationship are $a_{0}(\pi\eta, K\overline{K})$, $f_{0}(\pi\pi, K\overline{K})$ and $\sigma(\pi\pi)$.}
	\label{fig:figure1}
\end{figure}

%
%%%
%%%%%%%%%%%%%%%%% II. Theoretical frame and the wave function %%%%%%%%%%%%%%%%%%%%%%%%%%%%%%%%
%%%
%
\section{The Theoretical Framework and Helicity Amplitudes}\label{sec:pert}

\subsection{The Wave Functions}

The relevant weak effective Hamiltonian of the quasi-two-body $\overline{B}^{0}_{s} \rightarrow S[ \rightarrow P_{1}P_{2}]S$ decays can be written as~\cite{Buchalla:1995vs}

\begin{equation}
{\cal H}_{eff}=\frac{G_{F}}{\sqrt2}\big\{V_{ub}V^{*}_{us}[C_{1}(\mu)O_{1}(\mu)+C_{2}(\mu)O_{2}(\mu)]-V_{tb}V^{*}_{ts}[\sum^{10}_{i=3}C_{i}(\mu)O_{i}(\mu)]\big\},
\end{equation}
with the Fermi constant $G_{F}=1.66378\times10^{-5}$ GeV$^{-2}$, the local four-quark operator $O_{i}(\mu)$, the corresponding Wilson coefficient $C_{i}(\mu)$, and $V_{ub}V^{*}_{us}$ and $V_{tb}V^{*}_{ts}$ are Cabibbo-Kobayashi-Maskawa (CKM) factors. The Feynman diagrams involved in this work are illustrated in Fig.~\ref{fig:figure1}.

Based on the light-cone coordinates, we let the $\overline{B}^0_{s}$ meson stay at rest, and choose the $P_{1}P_{2}$ meson pair and the final-state $S$ movement along the direction of $n=(1,0,0_{T})$ and $v=(0,1,0_{T})$, respectively. So the $\overline{B}^0_{s}$ meson momentum $p_{B}$, the total momentum $p=p_{1}+p_{2}$ of the $P_{1}P_{2}$ meson pair, and the momentum $p_{3}$ of the final-state $S$ are considered as

\begin{equation}
\begin{split}
&p_{B}=\frac{M_{B}}{\sqrt{2}}(1,1,0_{T}),                 \\
&p=\frac{M_{B}}{\sqrt{2}}(1-r^2,\eta,0_{T}),              \\
&p_{3}=\frac{M_{B}}{\sqrt{2}}(r^2,1-\eta,0_{T}),
\end{split}
\end{equation}

where $M_{B}$ is the mass of $B^{0}_{s}$, $r=\frac{m_{S}}{M_{B}}$ is the mass ratio, $m_{S}$ refers to the mass of the final-state $S$. We think the variable $\eta=\omega^2/({M_{B}^2}-{m_{S}^2})$, and $\omega$ is the invariant mass of the $P_{1}P_{2}$ meson pair, which satisfies the relation $\omega^2=p^2$. Meanwhile, we define $\zeta=p^+_{1}/p^+$ as one of the $P_{1}P_{2}$ meson pair's momentum fractions. Accordingly, the kinematic variables of other components in the meson pair can be expressed as

\begin{equation}
\begin{split}
& p^-_{1}=\frac{M_{B}}{\sqrt{2}}(1-\zeta)\eta ,  \\
&p^+_{2}=\frac{M_{B}}{\sqrt{2}}(1-\zeta)(1-r^2),\\
&p^-_{2}=\frac{M_{B}}{\sqrt{2}}\zeta\eta .
\end{split}
\end{equation}

We adopt $x_{B}$, $z$, $x_{3}$ to indicate the momentum fraction of the light quark in each meson with the range from zero to unity. So the light quark's momentum of the $\overline{B}^0_{s}(k_{B})$, $P_{1}P_{2}(k)$ and $S(k_{3})$ are read as

\begin{equation}
\begin{split}
&k_{B}=(0,\frac{M_{B}}{\sqrt{2}}{x_{B}},k_{BT}),\\
&k=(\frac{M_{B}}{\sqrt{2}}z(1-r^2),0,k_{T}),\\
&k_{3}=(\frac{M_{B}}{\sqrt{2}}r^2x_{3},\frac{M_{B}}{\sqrt{2}}(1-\eta)x_{3},k_{3T}).
\end{split}
\end{equation}

In this work, the wave function of the hadron $B^{0}_{s}$ can be given by Refs.~\cite{C:2001oqa,Keum:2000wi,Keum:2000ph,HnLijhep2013}
\begin{equation}
\Phi_{B_{s}}=\frac{i}{\sqrt{2N_{c}}}(\not {p}_{B}+M_{B}){\gamma_{5}}{\phi_{B_{s}}({x_{B},b_{B}})},
\end{equation}
where $N_{c}=3$ is the color factor, and the distribution amplitude(DA) ${\phi_{B_{s}}({x_{B},b_{B}})}$ is expressed in the usual form, which is
\begin{equation}
\phi_{B_{s}}(x_{B},b_{B})=\emph{N}_{B}{{x_{B}}^2}(1-{x_{B}})^2\exp[-\frac{M^2_{B}{{x_{B}}^2}}{2\omega^2_{b}}-\frac{1}{2}(\omega_{b}{b_{B}})^2],
\end{equation}
the factor $\emph{N}_{B}$ can be calculated by the normalization $\int^{1}_{0}\phi_{B_{s}}(x_{B},b_{B}=0)\emph{dx}=\emph{f}_{B}/({2}{\sqrt{{2}{\emph{N}}_{c}}})$ with the $B^{0}_{s}$ meson decay constant $f_{B}$. And we select the shape parameter $\omega_{b}=0.50\pm0.05$ {\rm GeV}~\cite{Ali:2007ff}.

For the light scalar mesons $a_{0}(980)$ and $f_{0}(980)$, the wave function can be found in Refs.~\cite{Cheng:2006yov,H:2008da}

\begin{equation}
\Phi_{S}(x_{3})=\frac{1}{2\sqrt{2\emph{N}_{c}}}[\not{p}_{3}\phi_{S}(x_{3})+m_{S}\phi^{S}_{S}(x_{3})+m_{S}(\not{v}\not{n}-1)\phi^{T}_{S}(x_{3})]
\end{equation}
with the twist-2 distribution amplitude $\phi_{S}$, which can be expanded by the Gegenbauer polynomials~\cite{Cheng:2006yov,H:2008da}:
\begin{equation}
\phi_{S}(x_{3},\mu)=\frac{3}{\sqrt{2N_{c}}}x_{3}(1-x_{3})\left\{f_{S}(\mu)+\overline{f}_{S}(\mu)\sum^{\infty}_{m=1,3}B_{m}(\mu)C^{3/2}_{m}(2x_{3}-1)\right\},
\end{equation}
$\phi^{S}_{S}$ and $\phi^{T}_{S}$ are the twist-3 distribution amplitudes, their asymptotic forms can be written as:
\begin{equation}
\phi^{S}_{S}(x_{3},\mu)=\frac{1}{2\sqrt{2N_{c}}}\overline{f}_{S}(\mu),
\end{equation}
\begin{equation}
\phi^{S}_{S}(x_{3},\mu)=\frac{1}{2\sqrt{2N_{c}}}\overline{f}_{S}(\mu)(1-2x_{3}),
\end{equation}
where $B_{m}$ is the Gegenbauer moment, $C^{3/2}_{m}(2x_{3}-1)$ denotes the Gegenbauer polynomials, $f_{S}$ and $\overline{f}_{S}$ stand for the vector and scalar decay constants of the light scalar mesons $a_{0}$ and $f_{0}$, respectively. It is obvious that only the odd Gegenbauer moments are considered in the DA of $\phi_{S}$, the even Gegenbauer coefficients $B_{m}$ are suppressed because of the conservation of charge conjugation invariance or vector current, and we just notice the Gegenbauer moments $B_{1}$ and $B_{3}$ since the small contribution of higher order Gegenbauer moments can be ignored. On the basis of QCD sum rules with the default scale $\mu=1$ GeV, for the light scalar meson $a_{0}$~\cite{Cheng:2006yov,H:2008da}, we adopt the Gegenbauer moments $B_{1}=-0.93\pm0.10$ and $B_{3}=0.14\pm0.08$, and the scalar decay constant value of which can be taken as $\overline{f}_{a_{0}}=0.365\pm0.020$ . Meanwhile, the Gegenbauer moments and scalar decay constants of the scalar meson $f_{0}$ can be listed as ~\cite{Cheng:2006yov,H:2008da}:

\begin{equation}
\begin{split}
&\overline{f}_{S}=\overline{f}^{n}_{f_{0}}=\overline{f}^{s}_{f_{0}}=0.370\pm0.020\hspace{0.2cm} {\rm GeV},\\
&B^{n}_{1}=-0.78\pm0.08,     B^{n}_{3}=0.02\pm0.07,\\
&B^{s}_{1,3}=0.8B^{n}_{1,3}.
\end{split}
\end{equation}
Here, we take the same value of the two decay constants $\overline{f}^{n}_{f_{0}}$ and $\overline{f}^{s}_{f_{0}}$~\cite{Cheng:2006yov}. In this article, we select the vector decay constants $f_{S}=0$ due to the discussions in the Ref.~\cite{Z:2020aaa}.

For the distribution amplitudes of the final-state $P_{1}P_{2}$ meson pair, we adopt the consistent form of the S-wave pion pair\cite{C:2003vda,W:2015yrm,U:2014vb}:

\begin{equation}
\Phi^{S}_{P_{1}P_{2}}=\frac{1}{\sqrt{2N_{c}}}[\not{p}\phi_{S}(z,\zeta,\omega^{2})+\omega\phi^{s}_{S}(z,\zeta,\omega^{2})+\omega(\not{n}\not{v}-1)\phi^{t}_{S}(z,\zeta,\omega^{2})],
\end{equation}
where the leading-twist distribution amplitude $\phi_{S}$ and the twist-3 DAs $\phi^{s}_{S}$ and $\phi^{t}_{S}$ have similar expression forms as the corresponding twists of the light scalar meson obtained using the timelike form factor by replacing the original scalar decay constants. The asymptotic expression of the light-cone distribution amplitudes $\phi^{(s,t)}_{S}$ are as follow~\cite{R:2019ngy,W:2015yrm}:

\begin{equation}
\begin{split}
&\phi_{S}(z,\zeta,\omega^{2})=\frac{9F_{S}(\omega)}{\sqrt{2N_{c}}}a_{2}z(1-z)(1-2z),\\
&\phi^{s}_{S}(z,\zeta,\omega^{2})=\frac{F_{S}(\omega)}{2\sqrt{2N_{c}}},\\
&\phi^{t}_{S}(z,\zeta,\omega^{2})=\frac{F_{S}(\omega)}{2\sqrt{2N_{c}}}(1-2z).
\end{split}
\end{equation}
with the Gegenbauer moment $a_{2}=0.3\pm0.1$ for $a_{0}$ and $a_{2}=0.3\pm0.2$ for $f_{0}$\cite{Aaij:2013zpx,Ye:2019qu}. $F_{S}(\omega)$ is the timelike form factor, which can be described well in terms of the relative Breit-Wigner~\cite{Back:2018fqe}. However, for the $a_{0}$ (or $f_{0}$) resonance, the main decay channels are $\pi\eta(\pi\pi)$ and $K\overline{K}$. The relative Breit-Wigner line shape cannot be well adapted to the timelike form factor of $a_{0}$ and $f_{0}$, because both of the $a_{0}$ and $f_{0}$ resonances are very close to the $K\overline{K}$ threshold, which greatly influences the resonance shape. As such, we choose the widely accepted prescription proposed by Flatt$\acute{e}$~\cite{M:1976tpn}, which are given as~\cite{R:2019ngy}:

\begin{equation}
F^{a_{0}}_{S}(\omega)=\frac{C_{a_{0}}m^{2}_{a_{0}}}{m^{2}_{a_{0}}-\omega^{2}-i(g^{2}_{\pi\eta}\rho_{\pi\eta}+g^{2}_{KK}\rho_{KK})},
\end{equation}
for the $a_{0}$ resonance and

\begin{equation}
F^{f_{0}}_{S}(\omega)=\frac{m^{2}_{f_{0}}}{m^{2}_{f_{0}}-\omega^{2}-im_{f_{0}}(g_{\pi\pi}\rho_{\pi\pi}+g_{KK}\rho_{KK}F^{2}_{KK})},
\end{equation}
for the $f_{0}$ resonance~\cite{L:2021vb,Z:2021hao,Yan:2022mao}. In the case of the timelike form factor of the $a_{0}$ resonance, $C_{a_{0}}=|C_{a_{0}}|e^{i\phi_{a_{0}}}$ is the complex amplitude of the intermediate state $a_{0}$, with different values for the final states $\pi\eta$ and $K\overline{K}$. For the $K\overline{K}$ channel, the magnitude $|C^{KK}_{a_{0}}|=1.07$ and phase $\phi_{a_{0}}=82^{\circ}$~\cite{R:2016mao}. Then, the phase of the $\pi\eta$ system is consistent with that of the $K\overline{K}$ system, and the module of the magnitude satisfies the relation $C^{\pi\eta}_{a_{0}}/C^{KK}_{a_{0}}=g_{a_{0}\pi\eta}/g_{a_{0}KK}$ according to the discussions in Ref.~\cite{R:2019ngy}. The definition of the strong coupling constants $g_{a_{0}KK}(g_{a_{0}\pi\eta})$ can be found in the literature~\cite{J:2022wa,W:2020lbd}. In this article, we take the coupling constants as $g_{\pi\eta}=0.324$ GeV and $g^{2}_{KK}/g^{2}_{\pi\eta}=1.03$ through the Crystal Barrel experiment~\cite{Abele:1998hls}. The values of the constants $g_{a_{0}KK}$ and $g_{a_{0}\pi\eta}$ can be got with the relation $g_{KK}(g_{\pi\eta})=g_{a_{0}KK}(g_{a_{0}\pi\eta})/(4\sqrt{\pi})$. Meanwhile, we employ the coupling constants $g_{\pi\pi}=0.165\pm0.018$ GeV and $g_{KK}/g_{\pi\pi}=4.21\pm0.33$ for $f_{0}$~\cite{Back:2018fqe}, and introduce the factor $F_{KK}=e^{{-\alpha}q^{2}}$ into the timelike form factor $F^{f_{0}}_{S}(\omega)$ to suppress the $K\overline{K}$ contribution with the parameter $\alpha=2.0\pm1.0$ GeV$^{-2}$~\cite{R:2014uea}. In addition, the $\rho$ factors are chosen as:

\begin{equation}
\rho_{\pi\eta}=\sqrt{[1-(\frac{m_{\eta}-m_{\pi}}{\omega})^{2}][1-(\frac{m_{\eta}+m_{\pi}}{\omega})^{2}]},
\end{equation}

\begin{equation}
\rho_{\pi\pi}=\frac{2}{3}\sqrt{1-\frac{4m^{2}_{\pi^{\pm}}}{\omega^{2}}}+\frac{1}{3}\sqrt{1-\frac{4m^{2}_{\pi^{0}}}{\omega^{2}}},
\end{equation}

\begin{equation}
\rho_{KK}=\frac{1}{2}\sqrt{1-\frac{4m^{2}_{K^{\pm}}}{\omega^{2}}}+\frac{1}{2}\sqrt{1-\frac{4m^{2}_{K^{0}}}{\omega^{2}}}.
\end{equation}
The shape of $\sigma$ resonance can be well described by the Breit-Wigner model because it is a narrow intermediate resonance~\cite{R:2013az,Liang:2022xti}:

\begin{equation}
F^{\sigma}_{S}(\omega)=\frac{C_{\sigma}m^{2}_{\sigma}}{m^{2}_{\sigma}-\omega^{2}-im_{\sigma}\Gamma(\omega)},
\end{equation}
with the factor $C_{\sigma}=3.50$~\cite{R:2014uea}. The energy-dependent width $\Gamma(\omega)$ in the case of a scalar resonance decaying into pion pair can be parameterized as~\cite{W:2015yrm}:

\begin{equation}
\Gamma(\omega)=\Gamma_{0}\frac{m_{\sigma}}{\omega}(\frac{\omega^{2}-4m^{2}_{\pi}}{m^{2}_{\sigma}-4m^{2}_{\pi}})^{\frac{1}{2}},
\end{equation}
where $\Gamma_{0}=0.40$ GeV is the width of the resonance.

In this paper, we take the mixing relation for $f_{0}-\sigma$ system~\cite{A:2005wy},

\begin{equation}
\begin{pmatrix}
\sigma \\
f_{0}
\end{pmatrix}
=
\begin{pmatrix}
\cos\theta & -\sin\theta \\
\sin\theta & \cos\theta
\end{pmatrix}
\begin{pmatrix}
f_{n} \\
f_{s}
\end{pmatrix}
\end{equation}
with

\begin{equation}
f_{n}=\frac{1}{\sqrt{2}}(u\overline{u}+d\overline{d}),\\
f_{s}=s\overline{s}.
\end{equation}

\subsection{Helicity Amplitudes}

The $\overline{B}^{0}_{s} \rightarrow a_{0}[ \rightarrow \pi^{0}\eta, K\overline{K}]a_{0}$ decays only have annihilation Feynman diagrams, and their helicity amplitudes are written as

\begin{equation}
\begin{split}
{\cal A}(\overline{B}^{0}_{s} \rightarrow a^{+}_{0}[ \rightarrow \pi^{+}\eta, K^{+}\overline{K}^{0}]a^{-}_{0})=&\frac{G_{F}}{\sqrt{2}}V_{ub}V^{*}_{us}[(C_{1}+\frac{1}{3}C_{2})F^{'LL}_{e}+C_{2}(M^{'LL}_{g})]\\
&-V_{tb}V^{*}_{ts}[(a_{3}+a_{5}+a_{7}+a_{9})F^{'LL}_{e}\\
&+(a_{3}+a_{5}-\frac{1}{2}(a_{7}+a_{9}))F^{LL}_{e}\\
&+(C_{4}-\frac{1}{2}C_{10})M^{LL}_{g}+(C_{4}+C_{10})M^{'LL}_{g}\\
&+(C_{6}-\frac{1}{2}C_{8})M^{SP}_{g}+(C_{6}+C_{8})M^{'SP}_{g}],
\label{bsa0}
\end{split}
\end{equation}

\begin{equation}
\begin{split}
{\cal A}(\overline{B}^{0}_{s} \rightarrow a^{-}_{0}[ \rightarrow \pi^{-}\eta, K^{-}\overline{K}^{0}]a^{+}_{0})=&\frac{G_{F}}{\sqrt{2}}V_{ub}V^{*}_{us}[(C_{1}+\frac{1}{3}C_{2})F^{LL}_{e}+C_{2}(M^{LL}_{g})]\\
&-V_{tb}V^{*}_{ts}[(a_{3}+a_{5}-\frac{1}{2}(a_{7}+a_{9}))F^{'LL}_{e}\\
&+(a_{3}+a_{5}+a_{7}+a_{9})F^{LL}_{e}\\
&+(C_{4}+C_{10})M^{LL}_{g}+(C_{4}-\frac{1}{2}C_{10})M^{'LL}_{g}\\
&+(C_{6}+C_{8})M^{SP}_{g}+(C_{6}-\frac{1}{2}C_{8})M^{'SP}_{g}],
\end{split}
\end{equation}

\begin{equation}
\begin{split}
{\cal A}(\overline{B}^{0}_{s} \rightarrow a^{0}_{0}[ \rightarrow \pi^{0}\eta, K^{+}K^{-}]a^{0}_{0})=&\frac{G_{F}}{2}V_{ub}V^{*}_{us}[(C_{1}+\frac{1}{3}C_{2})(F^{LL}_{e}+F^{'LL}_{e})\\
&+C_{2}(M^{LL}_{g}+M^{'LL}_{g})]\\
&-V_{tb}V^{*}_{ts}[(2(a_{3}+a_{5})+\frac{1}{2}(a_{7}+a_{9}))(F^{LL}_{e}+F^{'LL}_{e})\\
&+(2C_{4}+\frac{1}{2}C_{10})(M^{LL}_{g}+M^{'LL}_{g})\\
&+(2C_{6}+\frac{1}{2}C_{8})(M^{SP}_{g}+M^{'SP}_{g})],
\end{split}
\end{equation}
where $F_{e}(F^{'}_{e})$ stands for the contributions of the factorizable annihilation diagrams shown in Fig.~\ref{fig:figure1}(e1),(f1)(e2),(f2), and $M_{g}(M^{'}_{g})$ comes from the nonfactorizable annihilation diagrams in Fig.~\ref{fig:figure1}(g1),(h1)(g2),(h2). In subsequent calculation, the function $F_{a}(F^{'}_{a})$ denotes the contributions of the factorizable emission diagrams in Fig.~\ref{fig:figure1}(a1),(b1)(a2),(b2), and $M_{c}(M^{'}_{c})$ comes from the nonfactorizable emission diagrams in Fig.~\ref{fig:figure1}(c1),(d1)(c2),(d2). The specific expressions for the aforementioned functions are presented in Appendix. The superscripts $LL$, $LR$, $SP$ represent the contributions of $(V-A)(V-A)$, $(V-A)(V+A)$ and $(S-P)(S+P)$ vertices, respectively.

Based on the $f_{0}-\sigma$ mixing scheme, the helicity amplitudes of the $\overline{B}^{0}_{s} \rightarrow f_{0}[ \rightarrow \pi^{+}\pi^{-}, K^{+}K^{-}]f_{0}$ and $\overline{B}^{0}_{s} \rightarrow \sigma[ \rightarrow \pi^{+}\pi^{-}]\sigma$ decays are given by

\begin{equation}
\begin{split}
{\cal A}(\overline{B}^{0}_{s} \rightarrow f_{0}[ \rightarrow \pi^{+}\pi^{-}, K^{+}K^{-}]f_{0})=&\sin^{2}\theta{\cal A}(\overline{B}^{0}_{s} \rightarrow f_{n}[ \rightarrow \pi^{+}\pi^{-}, K^{+}K^{-}]f_{n})\\
&+\frac{1}{2}\sin2\theta{\cal A}(\overline{B}^{0}_{s} \rightarrow f_{s}[ \rightarrow \pi^{+}\pi^{-}, K^{+}K^{-}]f_{n})\\
&+\frac{1}{2}\sin2\theta{\cal A}(\overline{B}^{0}_{s} \rightarrow f_{n}[ \rightarrow \pi^{+}\pi^{-}, K^{+}K^{-}]f_{s})\\
&+\cos^{2}\theta{\cal A}(\overline{B}^{0}_{s} \rightarrow f_{s}[ \rightarrow \pi^{+}\pi^{-}, K^{+}K^{-}]f_{s}),
\end{split}
\label{f0f0}
\end{equation}

\begin{equation}
\begin{split}
{\cal A}(\overline{B}^{0}_{s} \rightarrow \sigma[  \rightarrow \pi^{+}\pi^{-}]\sigma)=&\cos^{2}\theta{\cal A}(\overline{B}^{0}_{s} \rightarrow f_{n}[ \rightarrow \pi^{+}\pi^{-}]f_{n})\\
&-\frac{1}{2}\sin2\theta{\cal A}(\overline{B}^{0}_{s} \rightarrow f_{s}[ \rightarrow \pi^{+}\pi^{-}]f_{n})\\
&-\frac{1}{2}\sin2\theta{\cal A}(\overline{B}^{0}_{s} \rightarrow f_{n}[ \rightarrow \pi^{+}\pi^{-}]f_{s})\\
&+\sin^{2}\theta{\cal A}(\overline{B}^{0}_{s} \rightarrow f_{s}[ \rightarrow \pi^{+}\pi^{-}]f_{s}),
\end{split}
\label{sigma}
\end{equation}
with

\begin{equation}
\begin{split}
{\cal A}(\overline{B}^{0}_{s} \rightarrow f_{s}[ \rightarrow \pi^{+}\pi^{-}, K^{+}K^{-}]f_{s})=&-G_{F}V_{tb}V^{*}_{ts}[(a_{6}-\frac{1}{2}a_{8})(F^{SP}_{a}+F^{SP}_{e}+F^{'SP}_{a}+F^{'SP}_{e})\\
&+(a_{3}+a_{4}+a_{5}-\frac{1}{2}(a_{7}+a_{9}+a_{10}))(F^{LL}_{e}+F^{'LL}_{e})\\
&+(C_{3}+C_{4}-\frac{1}{2}(C_{9}+C_{10}))(M^{LL}_{c}+M^{LL}_{g}+M^{'LL}_{c}+M^{'LL}_{g})\\
&+(C_{5}-\frac{1}{2}C_{7})(M^{LR}_{c}+M^{LR}_{g}+M^{'LR}_{c}+M^{'LR}_{g})\\
&+(C_{6}-\frac{1}{2}C_{8})(M^{SP}_{c}+M^{SP}_{g}+M^{'SP}_{c}+M^{'SP}_{g})],
\end{split}
\end{equation}

\begin{equation}
\begin{split}
{\cal A}(\overline{B}^{0}_{s} \rightarrow f_{s}[ \rightarrow \pi^{+}\pi^{-}, K^{+}K^{-}]f_{n})=&\frac{G_{F}}{2}[V_{ub}V^{*}_{us}C_{2}M^{LL}_{c}\\
&-V_{tb}V^{*}_{ts}((2C_{4}+\frac{1}{2}C_{10})M^{LL}_{c}+(2C_{6}+\frac{1}{2}C_{8})M^{SP}_{c})].
\end{split}
\label{bsfn}
\end{equation}
Equation (\ref{bsfn}) is also applicable for $\overline{B}^{0}_{s} \rightarrow f_{n}[ \rightarrow \pi^{+}\pi^{-}, K^{+}K^{-}]f_{s}$ decay after replacing $M^{LL,SP}_{c}$ with $M^{'LL,'SP}_{c}$. Meanwhile, $\overline{B}^{0}_{s} \rightarrow f_{n}[ \rightarrow \pi^{+}\pi^{-}, K^{+}K^{-}]f_{n}$ decay has the same amplitude as $\overline{B}^{0}_{s} \rightarrow a^{0}_{0}[ \rightarrow \pi^{0}\eta, K^{+}K^{-}]a^{0}_{0}$ decay.

%%%--=================================================================
%%%=====           Numerical evaluation and discussions   ============
%%5===================================================================

\section{Numerical Results And Discussions}\label{sec:numer}

With the decay amplitudes $\cal A$, the differential branching ratio for the $\overline{B}^{0}_{s} \rightarrow S[ \rightarrow P_{1}P_{2}]S$ decays can be taken as

\begin{equation}
\frac{d\cal B}{d\omega}=\frac{\tau\omega\mid\vec{p}_{1}\mid\mid\vec{p}_{3}\mid}{32\pi^{3}M_{B}^{3}}\overline{\mid\cal A\mid}^{2},
\end{equation}
where $\tau$ is the $B$ meson lifetime, $\mid\vec{p}_{1}\mid$ and $\mid\vec{p}_{3}\mid$ respectively denote the magnitudes of momentum for one of the $P_{1}P_{2}$ meson pair and the scalar meson $S$

\begin{equation}
\begin{split}
\mid\vec{p}_{1}\mid=\frac{\lambda^{1/2}(\omega^{2}, m_{P_{1}}^{2},m_{P_{2}}^{2})}{2\omega},\\
\mid\vec{p}_{3}\mid=\frac{\lambda^{1/2}(M_{B}^{2}, m_{s}^{2},\omega^{2})}{2\omega}
\end{split}
\end{equation}
with the K\"{a}ll\'{e}n function $\lambda(a,b,c)=a^{2}+b^{2}+c^{2}-2(ab+ac+bc)$.

In Table \ref{tab}, we present the input parameters used in the calculations, including the masses of the mesons, the decay constant and lifetime of the $B^{0}_{s}$ meson, and the Wolfenstein parameters of the CKM matrix elements~\cite{P:2020mia,H:2005ira,X:2019rlo}.

\begin{table}[htbp]
\centering
\caption{Input parameters of the $\overline{B}^{0}_{s} \rightarrow S[ \rightarrow P_{1}P_{2}]S$ decays}
\label{tab}
\begin{tabular*}{\columnwidth}{@{\extracolsep{\fill}}llllll@{}}
\hline
\hline
\\
&$M_{B}=5.367$ {\rm GeV}       &$m_{b}=4.2$ {\rm GeV}     &$f_{B}=227.2\pm3.4$ {\rm MeV}    &$\tau=1.509$ {\rm ps}\\
                                  \\
&$m_{a_{0}}=0.98\pm0.02$ {\rm GeV}     &$m_{f_{0}}=0.99\pm0.02$ {\rm GeV}      &$m_{\sigma}=0.5$ {\rm GeV}     &$m_{f_{n}}=0.99$ {\rm GeV}\\
                                  \\
&$m_{f_{s}}=1.02$ {\rm GeV}       &$m_{f_{0}(1500)}=1.50$ {\rm GeV}    &$m_{\pi^{\pm}}=0.14$ {\rm GeV}      &$m_{\pi^{0}}=0.135$ {\rm GeV}\\
                                  \\
&$m_{K^{\pm}}=0.494$ {\rm GeV}     &$m_{K^{0}}=0.498$ {\rm GeV}    &$m_{\eta}=0.548$ {\rm GeV} \\
                                  \\
&$\lambda=0.22453\pm0.00044$  &$\emph{A}=0.836\pm0.015$   &$\bar{\rho}=0.122^{+0.018}_{-0.017}$   &$\bar{\eta}=0.355^{+0.012}_{-0.011}$\\
                                  \\
\hline
\hline
\end{tabular*}
\end{table}

By using the helicity amplitudes and the input parameters, we predict the $CP$-averaged branching fractions of the $\overline{B}^{0}_{s} \rightarrow S[ \rightarrow P_{1}P_{2}]S$ decays in the pQCD approach, and make some comments on the results. In Table \ref{a}, we present the branching fractions of the $\overline{B}^{0}_{s} \rightarrow a_{0}[ \rightarrow \pi\eta, K\overline{K}]a_{0}$ decays. There are still many uncertainties in our calculation results. As shown in the Table \ref{a}, we primarily consider four types of errors, namely the shape parameter of $B$ meson $\omega_{b}=0.50\pm0.05$ GeV, the Gegenbauer moment $a_{2}=0.3\pm0.1$ for the $K\overline{K}(\pi\eta)$ pair, the Gegenbauer moments $B_{1}=-0.93\pm0.10$ and $B_{3}=0.14\pm0.08$ for the scalar meson $a_{0}$, and the scalar decay constant $\overline{f}_{a_{0}}=0.365\pm0.020$ GeV. We have neglected the uncertainties caused by the mass of $a_{0}$ and the Wolfenstein parameters $\lambda, A, \rho, \eta$ because they are typically very small. We notice that the main uncertainties come from the Gegenbauer moment in the wave function of the $K\overline{K}(\pi\eta)$ pair, thus, we look forward to obtaining more accurate experimental data in the future to reduce such errors.

\begin{table}[htbp]
	\centering
	\caption{Branching ratios for the $\overline{B}^{0}_{s} \rightarrow a_{0}[ \rightarrow \pi\eta, K\overline{K}]a_{0}$ decays in the pQCD approach.}
	\label{a}
	\begin{tabular*}{\columnwidth}{@{\extracolsep{\fill}}llllll@{}}
		\hline
		\hline
		&Decay Modes         & ${\cal B}$     \\
		\hline
		\\
		&$\overline{B}^{0}_{s} \rightarrow a^{0}_{0}[ \rightarrow \pi^{0}\eta]a^{0}_{0}$  & $ 8.31^{+3.42}_{-2.26}(\omega_{b})^{+6.06}_{-4.35}(a_{2})^{+1.84}_{-0.61}(B)^{+0.93}_{-0.87}(\overline{f}_{a_{0}})\times10^{-8}$      \\
		\\
		&$\overline{B}^{0}_{s} \rightarrow a^{0}_{0}[ \rightarrow K^{+}K^{-}]a^{0}_{0}$  & $ 9.87^{+3.84}_{-2.61}(\omega_{b})^{+6.76}_{-4.89}(a_{2})^{+2.92}_{-1.34}(B)^{+1.07}_{-1.05}(\overline{f}_{a_{0}})\times10^{-9}$      \\
		\\
		&$\overline{B}^{0}_{s} \rightarrow a^{+}_{0}[ \rightarrow \pi^{+}\eta]a^{-}_{0}$  & $ 8.58^{+4.62}_{-3.35}(\omega_{b})^{+5.33}_{-4.05}(a_{2})^{+1.01}_{-0.93}(B)^{+0.97}_{-0.91}(\overline{f}_{a_{0}})\times10^{-7}$      \\
		\\
		&$\overline{B}^{0}_{s} \rightarrow a^{+}_{0}[ \rightarrow K^{+}\overline{K}^{0}]a^{-}_{0}$  & $ 1.76^{+0.95}_{-0.69}(\omega_{b})^{+1.06}_{-0.81}(a_{2})^{+0.20}_{-0.18}(B)^{+0.19}_{-0.18}(\overline{f}_{a_{0}})\times10^{-7}$      \\
		\\
		&$\overline{B}^{0}_{s} \rightarrow a^{-}_{0}[ \rightarrow \pi^{-}\eta]a^{+}_{0}$  & $ 8.76^{+4.66}_{-3.39}(\omega_{b})^{+5.32}_{-4.06}(a_{2})^{+1.11}_{-1.02}(B)^{+0.98}_{-0.93}(\overline{f}_{a_{0}})\times10^{-7}$      \\
		\\
		&$\overline{B}^{0}_{s} \rightarrow a^{-}_{0}[ \rightarrow K^{-}K^{0}]a^{+}_{0}$  & $ 1.81^{+0.98}_{-0.70}(\omega_{b})^{+1.06}_{-0.82}(a_{2})^{+0.23}_{-0.18}(B)^{+0.20}_{-0.19}(\overline{f}_{a_{0}})\times10^{-7}$      \\
		\hline
		\hline
	\end{tabular*}
\end{table}

Meanwhile, we can find the branching ratios of the $\overline{B}^{0}_{s} \rightarrow a_{0}[ \rightarrow \pi\eta]a_{0}$ decays are much larger than that of the $\overline{B}^{0}_{s} \rightarrow a_{0}[ \rightarrow K\overline{K}]a_{0}$ decays, which can be explained by the fact that the phase space for $K\overline{K}$ is suppressed. In the Ref.~\cite{J:2022wa}, the authors has predicted the branching ratios of the $B^{0}_{s} \rightarrow a_{0}[ \rightarrow \pi\eta, K\overline{K}]h$ ($h$ denote the pseudoscalar meson $\pi$ or $K$) decays in the pQCD approach and get the results as follows: ${\cal B}(B^{0}_{s} \rightarrow a^{0}_{0}[ \rightarrow K^{-}K^{+}]\pi^{0})=0.04^{+0.00+0.01}_{-0.01-0.01}\times10^{-6}$, ${\cal B}(B^{0}_{s} \rightarrow a^{0}_{0}[ \rightarrow \pi^{0}\eta]\pi^{0})=0.54^{+0.03+0.18}_{-0.05-0.10}\times10^{-6}$, these are comparable to our results because these decay modes have the same components in the quark model and only annihilation contributions. The branching ratio of $\overline{B}^{0}_{s} \rightarrow a^{0}_{0}[ \rightarrow \pi^{0}\eta]a^{0}_{0}$ decay is much smaller than that of the $B^{0}_{s} \rightarrow a^{0}_{0}[ \rightarrow \pi^{0}\eta]\pi^{0}$ decay obviously, and the branching ratios of $\overline{B}^{0}_{s} \rightarrow a^{0}_{0}[ \rightarrow K^{+}K^{-}]a^{0}_{0}$ decay and $B^{0}_{s} \rightarrow a^{0}_{0}[ \rightarrow K^{-}K^{+}]\pi^{0}$ decay exhibit the same relationship, the reason may be that the QCD dynamics of the final state mesons $a_{0}$ and $\pi^{0}$ are different. At the same time, the authors also concluded that the branching fractions of the $\pi\eta$ channel was $5$ times larger than that of the $K\overline{K}$ channel with the resonance $a_{0}(980)$ in the Ref.~\cite{J:2022wa}. Next, we will use our calculations to investigate the value of $\Gamma(a_{0} \rightarrow K^{+}K^{-})/\Gamma(a_{0} \rightarrow \pi^{0}\eta)$ with the narrow-width approximation.

When the narrow-width approximation is considered, the branching ratio of the quasi-two-body decay can be written as

\begin{equation}
{\cal B}(B \rightarrow M_{1}(R \rightarrow )M_{2}M_{3})\simeq{\cal B}(B \rightarrow M_{1}R)\times{\cal B}(R \rightarrow M_{2}M_{3}),
\end{equation}
with the resonance $R$. We can define a ratio ${\cal R}_{1}$ as

\begin{equation}
\begin{split}
{\cal R}_{1}&=\frac{{\Gamma}(a_{0} \rightarrow K^{+}K^{-})}{{\Gamma}(a_{0} \rightarrow \pi^{0}\eta)}   \\
&=\frac{{\cal B}(\overline{B}^{0}_{s} \rightarrow a^{0}_{0}a^{0}_{0})\times{\cal B}(a^{0}_{0} \rightarrow K^{+}K^{-})}{{\cal B}(\overline{B}^{0}_{s} \rightarrow a^{0}_{0}a^{0}_{0})\times{\cal B}(a^{0}_{0} \rightarrow \pi^{0}\eta)}\\
&\simeq\frac{{\cal B}(\overline{B}^{0}_{s} \rightarrow a^{0}_{0}[ \rightarrow K^{+}K^{-}]a^{0}_{0})}{{\cal B}(\overline{B}^{0}_{s} \rightarrow a^{0}_{0}[ \rightarrow \pi^{0}\eta]a^{0}_{0})}\approx0.12.
\end{split}
\end{equation}
After considering the isospin relation $\Gamma(a_{0} \rightarrow K^{+}K^{-})=\Gamma(a_{0} \rightarrow K\overline{K})/2$, we obtain the relative partial decay width ${\Gamma}(a_{0} \rightarrow K\overline{K})/{\Gamma}(a_{0} \rightarrow \pi\eta)\approx0.24$. The OBELIX Collaboration acquired $\Gamma(a_{0} \rightarrow K\overline{K})/\Gamma(a_{0} \rightarrow \pi^{0}\eta)=0.57\pm0.16$ through the coupled channel analysis of $\pi^{+}\pi^{-}\pi^{0}$, $K^{+}K^{-}\pi^{0}$ and $K^{\pm}K^{0}_{S}\pi^{\mp}$~\cite{M:2003bgg}. In Ref.~\cite{Abele:1998hls}, the authors calculated the branching ratio of the $p\overline{p} \rightarrow a_{0}(980)\pi \rightarrow K\overline{K}\pi$ decay($(5.92^{+0.46}_{-1.01})\times10^{-4}$), and combined with the data ${\cal B}(p\overline{p} \rightarrow a_{0}(980)\pi; a_{0} \rightarrow \pi\eta)=(2.61\pm0.48)\times10^{-3}$ in the annihilation channel $\pi^{0}\pi^{0}\eta$~\cite{C:1994yn}, the ratio of the partial widths was determined as ${\Gamma}(a_{0} \rightarrow K\overline{K})/{\Gamma}(a_{0} \rightarrow \pi\eta)=0.23\pm0.05$. The WA102 Collaboration gained ${\Gamma}(a_{0} \rightarrow K\overline{K})/{\Gamma}(a_{0} \rightarrow \pi\eta)=0.166\pm0.01\pm0.02$ from the decay of $f_{1}(1285)$~\cite{Barberis:1998zzb}. The average relative partial decay widths is ${\Gamma}(a_{0} \rightarrow K\overline{K})/{\Gamma}(a_{0} \rightarrow \pi\eta)=0.183\pm0.024$ given by the Particle Data Group~\cite{P:2020mia}. It is obvious that our results are slightly lager than the average ratio, but are in agreement with the data in Ref.~\cite{Abele:1998hls} within errors.

\begin{figure}[htbp]
\centering
\begin{tabular}{l}
	\includegraphics[width=0.5\textwidth]{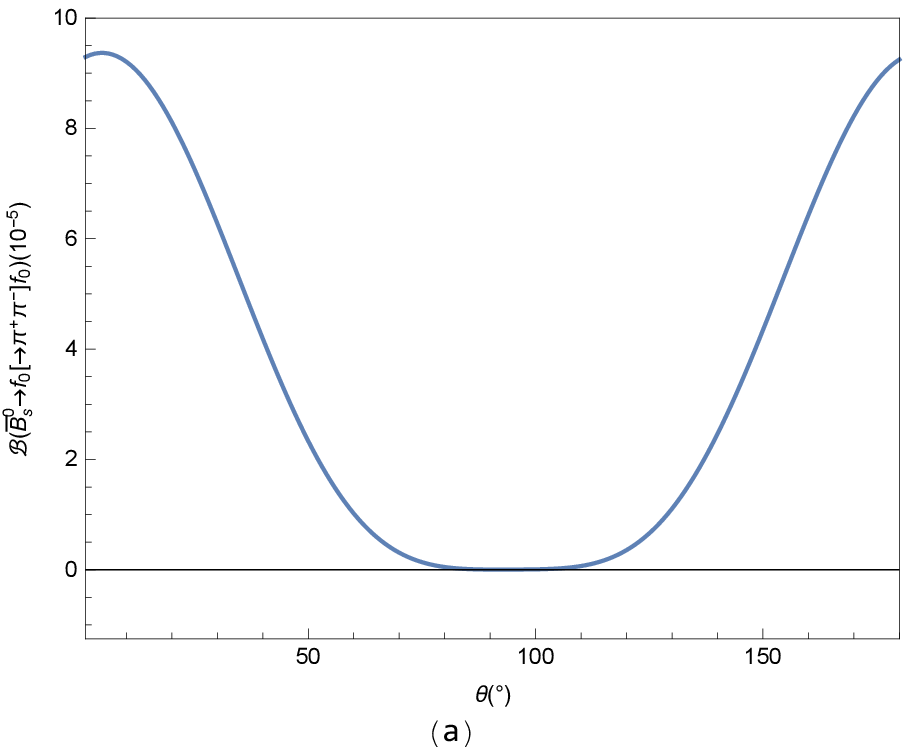}
	\includegraphics[width=0.5\textwidth]{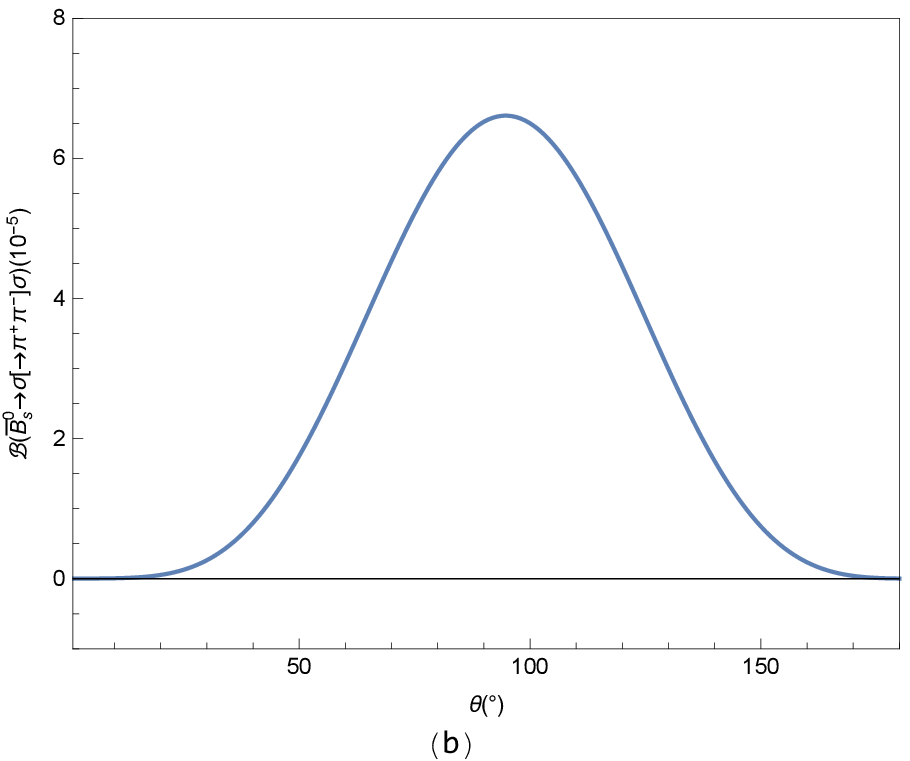}
	\end{tabular}
	\caption {(a)The branching ratio of $\overline{B}^{0}_{s} \rightarrow f_{0}[ \rightarrow \pi^{+}\pi^{-}]f_{0}$ decay on the mixing angle $\theta$;(b)The branching ratio of $\overline{B}^{0}_{s} \rightarrow \sigma[ \rightarrow \pi^{+}\pi^{-}]\sigma$ decay on the mixing angle $\theta$.}
\label{fig2}
\end{figure}

For the $\overline{B}^{0}_{s} \rightarrow f_{0}[ \rightarrow \pi^{+}\pi^{-},K^{+}K^{-}]f_{0}$ and $\overline{B}^{0}_{s} \rightarrow \sigma[ \rightarrow \pi^{+}\pi^{-}]\sigma$ decays, the scalars $f_{0}$ and $\sigma$ are not only regarded as $s\overline{s}$, but also contain $(u\overline{u}+d\overline{d})/\sqrt{2}$ in the quark model. The mixing angle $\theta$ is introduced into the $f_{0}-\sigma$ mixing mechanism. In this case, the decay amplitudes of the $\overline{B}^{0}_{s} \rightarrow f_{0}[ \rightarrow \pi^{+}\pi^{-},K^{+}K^{-}]f_{0}$ and $\overline{B}^{0}_{s} \rightarrow \sigma[ \rightarrow \pi^{+}\pi^{-}]\sigma$ decays  consist of four parts, and the total amplitudes are related to these four parts by the mixing angle $\theta$, which can be found in  Eqs.~(\ref{f0f0}) and ~(\ref{sigma}). We then plot the branching fractions of $\overline{B}^{0}_{s} \rightarrow f_{0}[ \rightarrow \pi^{+}\pi^{-}]f_{0}$ and $\overline{B}^{0}_{s} \rightarrow \sigma[ \rightarrow \pi^{+}\pi^{-}]\sigma$ decays dependent on the free parameter $\theta$ in Fig.~\ref{fig2}. The branching ratio of the $\overline{B}^{0}_{s} \rightarrow f_{0}[ \rightarrow \pi^{+}\pi^{-}]f_{0}$ decay obeys the cos law, as can be seen in Fig.~\ref{fig2}(a), while the other channel's contribution satisfies the sin law, as can be seen in Fig.~\ref{fig2}(b). Considering both decays, the range of the mixing angle $\theta$ can be set as [15$^{\circ}$, 82$^{\circ}$] and [105$^{\circ}$, 171$^{\circ}$] because the branching ratios of both decays are close to zero when $\theta$ takes other values. This range is larger than that of the two-body decays $\overline{B}^{0}_{s} \rightarrow f_{0}f_{0}$ and $\overline{B}^{0}_{s} \rightarrow \sigma\sigma$, but still consistent with the data in the Refs.~\cite{H:2003dk,Anisovich:2001uv,A:2005wy}.

The mixing angle $\theta$ is not fixed for the $f_{0}-\sigma$ system, and the value may vary in different works depending on the study requirements~\cite{L:2021vb,Wang:2019yu,Q:2019nm}. In the current study, we take $\theta=30^{\circ}$ to make numerical calculation, and this value also satisfies the results presented by the LHCb Collaboration~\cite{R:2013az}. Taking the mixing angle into account, the averaged branching ratios for the $\overline{B}^{0}_{s} \rightarrow f_{0}[ \rightarrow \pi^{+}\pi^{-},K^{+}K^{-}]f_{0}$ and $\overline{B}^{0}_{s} \rightarrow \sigma[ \rightarrow \pi^{+}\pi^{-}]\sigma$ decays are presented in Table \ref{f}.

\begin{table}[htbp]
	\centering
	\caption{Branching ratios for the $\overline{B}^{0}_{s} \rightarrow f_{0}[ \rightarrow \pi^{+}\pi^{-},K^{+}K^{-}]f_{0}$ and $\overline{B}^{0}_{s} \rightarrow \sigma[ \rightarrow \pi^{+}\pi^{-}]\sigma$ decays in the pQCD approach with the $f_{0}-\sigma$ mixing angle $\theta=30^{\circ}$.}
	\label{f}
	\begin{tabular*}{\columnwidth}{@{\extracolsep{\fill}}lllll@{}}
		\hline
		\hline
        &Decay Modes         & $\theta=30^{\circ}$     \\
		\hline
		\\
		&$\overline{B}^{0}_{s} \rightarrow f_{0}[ \rightarrow \pi^ {+}\pi^{-}]f_{0}$  & $ 6.07^{+2.04}_{-1.69}(\omega_{b})^{+0.91}_{-0.83}(a_{2})^{+0.70}_{-0.63}(B)^{+0.67}_{-0.64}(\overline{f}_{s})\times10^{-5}$      \\
		\\
		&$\overline{B}^{0}_{s} \rightarrow f_{0}[ \rightarrow K^ {+}K^{-}]f_{0}$      & $ 1.02^{+0.36}_{-0.29}(\omega_{b})^{+0.13}_{-0.12}(a_{2})^{+0.11}_{-0.09}(B)^{+0.11}_{-0.10}(\overline{f}_{s})\times10^{-5}$      \\
		\\
		&$\overline{B}^{0}_{s} \rightarrow f_{0}[ \rightarrow \pi^ {0}\pi^{0}]f_{0}$  & $ 3.03^{+1.02}_{-0.85}(\omega_{b})^{+0.45}_{-0.41}(a_{2})^{+0.35}_{-0.31}(B)^{+0.34}_{-0.32}(\overline{f}_{s})\times10^{-5}$      \\
		\\
		&$\overline{B}^{0}_{s} \rightarrow \sigma[ \rightarrow \pi^ {+}\pi^{-}]\sigma$& $ 3.01^{+0.85}_{-0.72}(\omega_{b})^{+0.36}_{-0.20}(a_{2})^{+0.09}_{-0.05}(B)^{+0.33}_{-0.31}(\overline{f}_{s})\times10^{-6}$      \\
		\\
		&$\overline{B}^{0}_{s} \rightarrow \sigma[ \rightarrow \pi^ {0}\pi^{0}]\sigma$& $ 1.50^{+0.42}_{-0.36}(\omega_{b})^{+0.18}_{-0.10}(a_{2})^{+0.05}_{-0.02}(B)^{+0.17}_{-0.16}(\overline{f}_{s})\times10^{-6}$      \\
		\hline
		\hline
	\end{tabular*}
\end{table}

As can be seen in Table \ref{f}, the major uncertainties originate from the shape parameter $\omega_{b}$ of the $B$ meson's wave function for both $\overline{B}^{0}_{s} \rightarrow f_{0}[ \rightarrow \pi^{+}\pi^{-},K^{+}K^{-}]f_{0}$ and $\overline{B}^{0}_{s} \rightarrow \sigma[ \rightarrow \pi^{+}\pi^{-}]\sigma$ decays. We still can find the branching ratio for the $\overline{B}^{0}_{s} \rightarrow f_{0}[ \rightarrow \pi^{+}\pi^{-}]f_{0}$ decay is lager than that of the $\overline{B}^{0}_{s} \rightarrow \sigma[ \rightarrow \pi^{+}\pi^{-}]\sigma$ decay by one order of magnitude, we think the reason for this result may be that the scalar meson $f_{0}$ has a greater mass than $\sigma$. When we take the mixing angle $\theta=0^{\circ}$, $f_{0}$ is regarded as the pure $s\overline{s}$, and the branching ratio of $\overline{B}^{0}_{s} \rightarrow f_{0}[ \rightarrow \pi^{+}\pi^{-}]f_{0}$ is about $9.29\times10^{-5}$. Whereas for $\overline{B}^{0}_{s} \rightarrow \sigma[ \rightarrow \pi^{+}\pi^{-}]\sigma$ decay, the branching ratio is very small with $\theta=0^{\circ}$, and the value will increase a lot after considering the mixing of $s\overline{s}$. Therefore, $\overline{B}^{0}_{s} \rightarrow f_{s}[ \rightarrow \pi^{+}\pi^{-}, K^{+}K^{-}]f_{s}$ makes the dominant contribution in the branching ratios. Numerical results of $\overline{B}^{0}_{s} \rightarrow f_{0}[ \rightarrow \pi^ {0}\pi^{0}]f_{0}$ and $\overline{B}^{0}_{s} \rightarrow \sigma[ \rightarrow \pi^ {0}\pi^{0}]\sigma$ decays are obtained with the isospin relationship ${\cal B}(f_{0}(\sigma) \rightarrow \pi^{+}\pi^{-})/{\cal B}(f_{0}(\sigma) \rightarrow \pi^{0}\pi^{0})=2$. The magnitudes of the predicted branching ratios are at the order of $10^{-5}\sim10^{-6}$, we expect these results can be tested by the LHCb and Belle II experiments in the nearly future.

To compare with existing data and further discuss our calculations, we then use the narrow-width approximation to study the $\overline{B}^{0}_{s} \rightarrow f_{0}[ \rightarrow \pi^{+}\pi^{-}, K^{+}K^{-}]f_{0}$ decays. The $\overline{B}^{0}_{s} \rightarrow f_{0}[ \rightarrow \pi^{+}\pi^{-}]f_{0}$ and $\overline{B}^{0}_{s} \rightarrow f_{0}[ \rightarrow K^{+}K^{-}]f_{0}$ decays have the same resonance, we can define a ratio ${\cal R}_{2}$ to describe the relationship between $f_{0} \rightarrow \pi^{+}\pi^{-}$ and $f_{0} \rightarrow K^{+}K^{-}$, which can be given as

\begin{equation}
\begin{split}
{\cal R}_{2}&=\frac{{\cal B}(f_{0} \rightarrow K^{+}K^{-})}{{\cal B}(f_{0} \rightarrow \pi^{+}\pi^{-})}   \\
&=\frac{{\cal B}(\overline{B}^{0}_{s} \rightarrow f_{0}f_{0})\times{\cal B}(f_{0} \rightarrow K^{+}K^{-})}{{\cal B}(\overline{B}^{0}_{s} \rightarrow f_{0}f_{0})\times{\cal B}(f_{0} \rightarrow \pi^{+}\pi^{-})}\\
&\simeq\frac{{\cal B}(\overline{B}^{0}_{s} \rightarrow f_{0}[ \rightarrow K^{+}K^{-}]f_{0})}{{\cal B}(\overline{B}^{0}_{s} \rightarrow f_{0}[ \rightarrow \pi^{+}\pi^{-}]f_{0})}\approx0.17,
\end{split}
\end{equation}
this ratio can be used to estimate the branching ratios for the $f_{0} \rightarrow \pi^{+}\pi^{-}$ and $f_{0} \rightarrow K^{+}K^{-}$ decays by using the formulas ${\cal B}(f_{0} \rightarrow \pi^{+}\pi^{-})=\frac{2}{4{\cal R}_{2}+3}$ and ${\cal B}(f_{0} \rightarrow K^{+}K^{-})=\frac{2{\cal R}_{2}}{4{\cal R}_{2}+3}$~\cite{Fleischer:2011hz}. So we can get

\begin{equation}
\begin{split}
{\cal B}(f_{0} \rightarrow \pi^{+}\pi^{-})\approx0.54,\\
{\cal B}(f_{0} \rightarrow K^{+}K^{-})\approx0.09.
\end{split}
\end{equation}
The BES Collaboration gained the relative branching ratios through the $\psi(2S) \rightarrow \gamma\chi_{c0}$ decays, where $\chi_{c0} \rightarrow f_{0}f_{0} \rightarrow \pi^{+}\pi^{-}\pi^{+}\pi^{-}$ or $\chi_{c0} \rightarrow f_{0}f_{0} \rightarrow \pi^{+}\pi^{-}K^{+}K^{-}$~\cite{Ablikim:2004vu,Ablikim:2005rhr}. Meanwhile, CLEO Collaboration has obtained ${\cal B}(f_{0} \rightarrow K^{+}K^{-})/{\cal B}(f_{0} \rightarrow \pi^{+}\pi^{-})=(25^{+17}_{-11})$\% and extracted ${\cal B}(f_{0} \rightarrow \pi^{+}\pi^{-})=(50^{+7}_{-9})$\% by using the results of BES Collaboration ~\cite{Ecklund:2009dvs}, and our calculations are still consistent with that in CLEO Collaboration.

\begin{figure}[htbp]
	\centering
	\begin{tabular}{l}
		\includegraphics[width=0.5\textwidth]{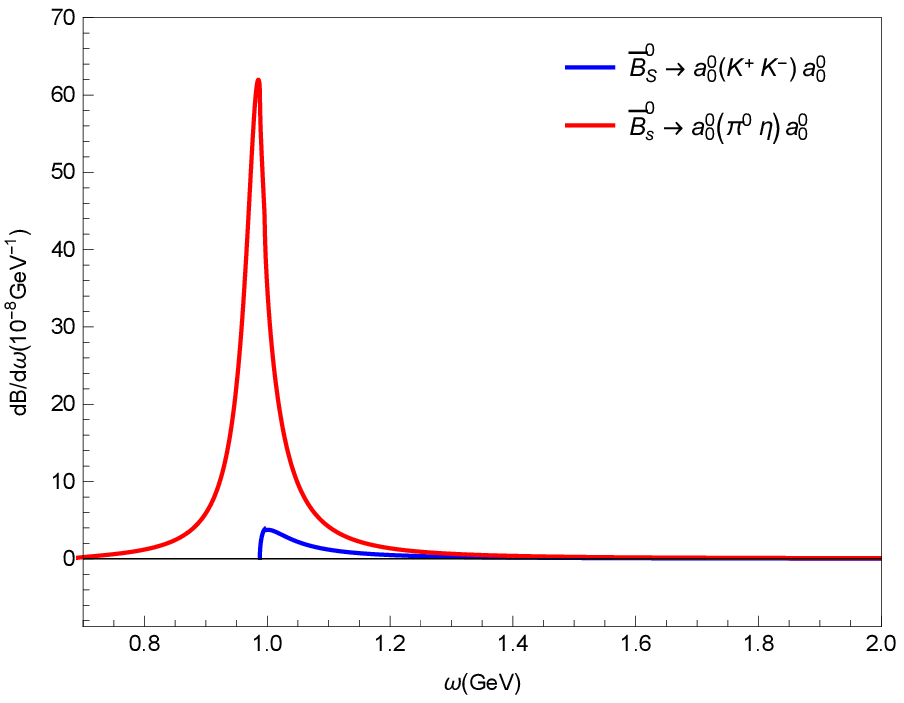}
		\includegraphics[width=0.5\textwidth]{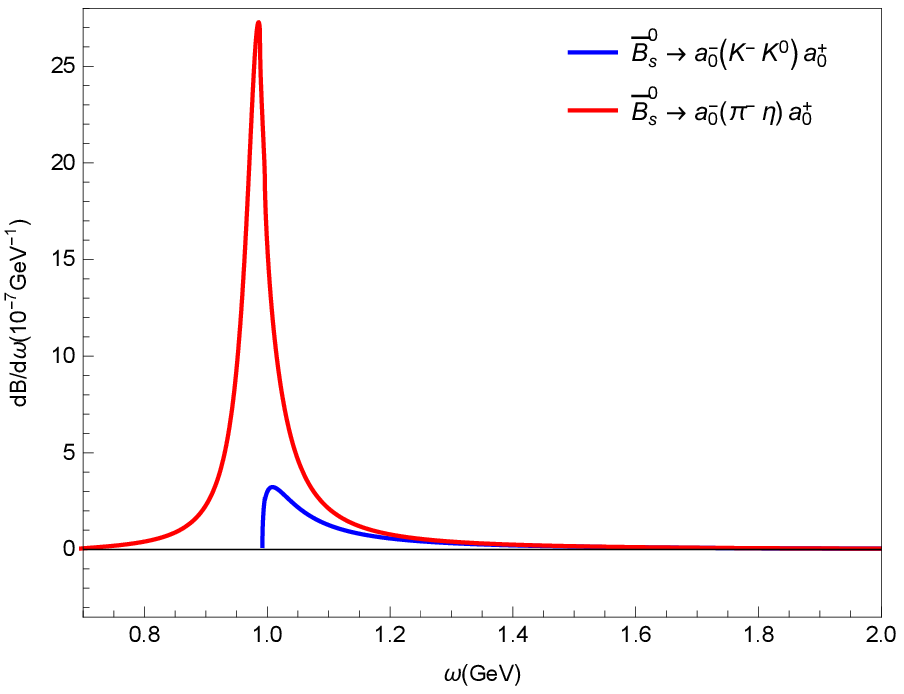}
	\end{tabular}
	\caption {Differential branching fractions of the $\overline{B}^{0}_{s} \rightarrow a_{0}[ \rightarrow K\overline{K}, \pi\eta]a_{0}$ decays.}
	\label{fig3}
\end{figure}

 \begin{figure}[htbp]
	\centering
	\begin{tabular}{l}
		\includegraphics[width=0.5\textwidth]{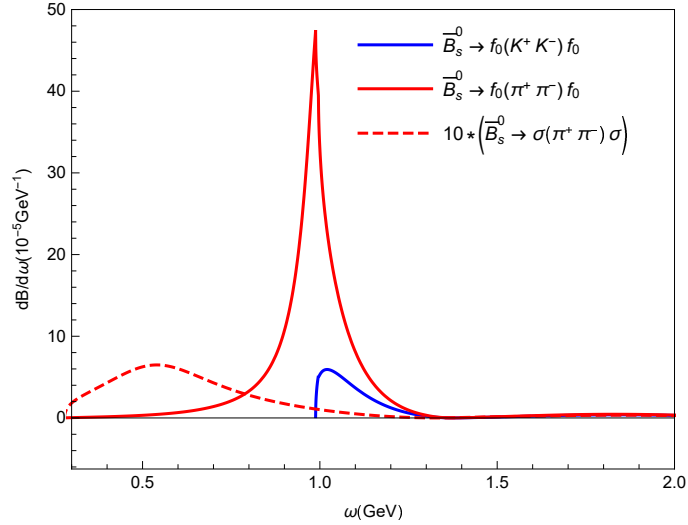}
	\end{tabular}
	\caption {Differential branching fractions of the $\overline{B}^{0}_{s} \rightarrow f_{0}[ \rightarrow \pi^{+}\pi^{-},K^{+}K^{-}]f_{0}$ and $\overline{B}^{0}_{s} \rightarrow \sigma[ \rightarrow \pi^{+}\pi^{-}]\sigma$ decays.}
	\label{fig4}
\end{figure}

In Fig.~\ref{fig3}, we graph the differential branching ratios of the $\overline{B}^{0}_{s} \rightarrow a_{0}[ \rightarrow  K\overline{K}, \pi\eta]a_{0}$ decays on the invariant mass, the results of $\overline{B}^{0}_{s} \rightarrow a^{0}_{0}[ \rightarrow K^{+}K^{-}, \pi^{0}\eta]a^{0}_{0}$ models are shown on the left and those of $\overline{B}^{0}_{s} \rightarrow a^{-}_{0}[ \rightarrow K^{-}K^{0}, \pi^{-}\eta]a^{+}_{0}$ models are shown on the right. The differential branching ratios of $\overline{B}^{0}_{s} \rightarrow f_{0}[ \rightarrow \pi^{+}\pi^{-}]f_{0}$, $\overline{B}^{0}_{s} \rightarrow f_{0}[ \rightarrow K^{+}K^{-}]f_{0}$, and $\overline{B}^{0}_{s} \rightarrow \sigma[ \rightarrow \pi^{+}\pi^{-}]\sigma$ decays on the $\pi\pi$ or $K\overline{K}$ invariant mass $\omega$ are presented in Fig.~\ref{fig4} with red solid line, blue solid line and red dashed line, respectively. For the $a_{0}$ resonance, the contributions of the $K\overline{K}$ channel are much smaller than that of the $\pi\eta$ channel, and for the $d{\cal B}(\overline{B}^{0}_{s} \rightarrow \sigma[ \rightarrow \pi^{+}\pi^{-}]\sigma)/d\omega$ mode, we magnify the results tenfold for easy viewing. From the figures, it is clear that the peak occurs around the resonance peak mass, and  the majority of the branching ratios are concentrated around the resonance state, basically in the range of $[m_{S}-\Gamma_{S}, m_{S}+\Gamma_{S}]$. Here, we don't plot the contributions of $\overline{B}^{0}_{s} \rightarrow a^{+}_{0}[ \rightarrow K^{+}\overline{K}^{0}, \pi^{+}\eta]a^{-}_{0}$ channels separately, because their results are very similar to the results of $\overline{B}^{0}_{s} \rightarrow a^{-}_{0}[ \rightarrow K^{-}K^{0}, \pi^{-}\eta]a^{+}_{0}$ channels.

\section{Summary} \label{sec:summary}

In this article, we predict the branching fractions of $\overline{B}^{0}_{s} \rightarrow a_{0}[ \rightarrow  K\overline{K}, \pi\eta]a_{0}$, $\overline{B}^{0}_{s} \rightarrow f_{0}[ \rightarrow \pi^{+}\pi^{-}, K^{+}K^{-}]f_{0}$ and $\overline{B}^{0}_{s} \rightarrow \sigma[ \rightarrow \pi^{+}\pi^{-}]\sigma$ decays with the pQCD approach firstly, where the scalars are considered as the $q\overline{q}$ state in the first scenario. Our results show that: (1)For the $\overline{B}^{0}_{s} \rightarrow a_{0}[ \rightarrow  K\overline{K}, \pi\eta]a_{0}$ decays, the largest branching ratio is ${\cal B}(\overline{B}^{0}_{s} \rightarrow a^{-}_{0}[ \rightarrow \pi^{-}\eta]a^{+}_{0})=8.76\times10^{-7}$, which is highly likely to be verified experimentally; (2)For the $\overline{B}^{0}_{s} \rightarrow f_{0}[ \rightarrow \pi^{+}\pi^{-}, K^{+}K^{-}]f_{0}$ and $\overline{B}^{0}_{s} \rightarrow \sigma[ \rightarrow \pi^{+}\pi^{-}]\sigma$ decays, we consider that the scalars $f_{0}$ and $\sigma$ contain $s\overline{s}$ and $(u\overline{u}+d\overline{d})/\sqrt{2}$ components in the quark model, so our calculations are carried out with the $f_{0}-\sigma$ mixing scheme, when the value of the mixing angle $\theta$ is taken as 30$^\circ$,  our results are at the order of $10^{-6}\sim10^{-5}$. Using the narrow-width approximation, we calculate the relative partial decay widths ${\Gamma}(a_{0} \rightarrow K\overline{K})/{\Gamma}(a_{0} \rightarrow \pi\eta)$ and the ratio ${\cal B}(f_{0} \rightarrow K^{+}K^{-})/{\cal B}(f_{0} \rightarrow \pi^{+}\pi^{-})$, which are in agreement with the existing experimental values. Our work has positive implications for understanding the QCD behavior of the scalars, and we also expect that our calculations can be tested by the LHCb and Belle II experiments in the future.

%%%--=================================================================
%%%=====            Acknowledgements        ==========================
%%5===================================================================

\section*{acknowledgments}

The authors would to thank Li Xin for some valuable discussions. This work is supported by the National Natural Science Foundation of China under Grant No.11047028.
%%%--=================================================================
%%%=====            Appendix       ==========================
%%5===================================================================
\section*{Appendix : FACTORIZATION FORMULAE} \label{sec:appendix}
%\numberwithin{equation}{section}
\appendix
\setcounter{equation}{0}
\renewcommand\theequation{A.\arabic{equation}}
In this Section, we list the factorization formulas that are used in Eqs.~(\ref{bsa0})-~(\ref{bsfn}). Acorrding to previous studies~\cite{Z:2021hao,Liang:2019xq}, we can find the formulas for the factorization diagrams related to the decays $B \rightarrow S[\rightarrow PP]V$. In our work, the formulas for the factorization diagrams for the case when the final-state scalars are emitted are roughly consistent with those shown in Ref.~\cite{Liang:2019xq}. In contrast, for the factorization diagrams where the $PP$ meson pair is emitted or for the annihilation diagrams, the formulas are different from that in the existing works because of the difference in the wave functions between the final-state scalars and other mesons. So we recalculate the contributions of the Feynman diagrams. First, we give the contribution of the factorization diagrams in Fig.~\ref{fig:figure1}(a1) and (b1) with different currents, which are

(1) $(V-A)(V-A)$

\begin{eqnarray}
\begin{split}
F^{LL}_{a}=&-4\pi C_{F}f_{s}M^{4}_{B}\int_{0}^{1}dx_{B}dz\int_{0}^{\infty}b_{B}db_{B}bdb\phi_{B}(x_{B},b_{B})\\
&\left\{[-2\overline{\eta}z\phi_{S}(z)+ \sqrt{(1-r^{2})\eta}\overline{\eta}(2r_{b}+r^{2}(z-2)-z)(\phi^{s}_{S}(z)-\phi^{t}_{S}(z))\right.\\
&\left.+4\sqrt{(1-r^{2})\eta}r_{b}r^{2}\phi^{t}_{S}(z)]\alpha_{s}(t_{a1})h_{a1}(\alpha_{a1}, \beta_{a1}, b_{B}, b)\exp[-S_{B}(t_{a1})-S(t_{a1})]S_{t}(z)\right.\\
&\left.+[2(r^{2}(x_{B}-\eta)-\eta\overline{\eta})\phi_{S}(z)-4\sqrt{(1-r^{2})\eta}\overline{\eta}\phi^{s}_{S}(z)]\alpha_{s}(t_{b1})h_{b1}(\alpha_{b1}, \beta_{b1}, b, b_{B})\right.\\
&\left.\times\exp[-S_{B}(t_{b1})-S(t_{b1})]S_{t}(\left|x_{B}-\eta\right|)\right\}
\end{split}
\end{eqnarray}

(2) $(V-A)(V+A)$

\begin{eqnarray}
	F^{LR}_{a}=F^{LL}_{a}
\end{eqnarray}

(3) $(S-P)(S+P)$

\begin{eqnarray}
\begin{split}
F^{SP}_{a}=&8\pi C_{F}\overline{f}_{s}M^{4}_{B}r\int_{0}^{1}dx_{B}dz\int_{0}^{\infty}b_{B}db_{B}bdb\phi_{B}(x_{B},b_{B})\\
&\left\{[(2r_{b}(1-r^{2}+\eta)+4(\eta\overline{z}-1))\phi_{S}(z)-2\sqrt{(1-r^{2})\eta}(z+\overline{\eta})(\phi^{s}_{S}(z)-\phi^{t}_{S}(z))-4\sqrt{(1-r^{2})\eta}\right.\\
&\left.\times(r^{2}\overline{z}+z)\phi^{t}_{S}(z)+8\sqrt{(1-r^{2})\eta}r_{b}\phi^{s}_{S}(z)]\alpha_{s}(t_{a1})h_{a1}(\alpha_{a1}, \beta_{a1}, b_{B}, b)\exp[-S_{B}(t_{a1})-S(t_{a1})]S_{t}(z)\right.\\
&\left.+[2(r^{2}-1)(x_{B}-2\eta)\phi_{S}(z)-4\sqrt{(1-r^{2})\eta}(r^{2}-\overline{x}_{B}-\eta)\phi^{s}_{S}(z)]\right.\\
&\left.\times\alpha_{s}(t_{b1})h_{b1}(\alpha_{b1}, \beta_{b1}, b, b_{B})\exp[-S_{B}(t_{b1})-S(t_{b1})]S_{t}(\left|x_{B}-\eta\right|)\right\}
\end{split}
\end{eqnarray}
with $\overline{x}_{i}=1-x_{i}, \overline{\eta}=1-\eta$. $C_{F}=\frac{4}{3}$ stands for the color factor. We take $F^{LL}_{a}=F^{LR}_{a}=0$ because of the vector decay constant's small value. And the nonfactorization diagrams in Fig.~\ref{fig:figure1} (c1) and (d1) give

(1) $(V-A)(V-A)$

\begin{eqnarray}
\begin{split}
M^{LL}_{c}=&\frac{4\pi C_{F}M^{4}_{B}}{\sqrt{2N_{c}}}\int_{0}^{1}dx_{B}dzdx_{3}\int_{0}^{\infty}b_{B}db_{B}b_{3}db_{3}\phi_{B}(x_{B},b_{B})\phi_{S}(x_{3})\\
&\left\{[4\overline{\eta}(x_{B}+(z-2)\eta+x_{3}\overline{\eta})\phi_{S}(z)+4\sqrt{(1-r^{2})\eta}\bar{z}\bar{\eta} (\phi^{s}_{S}(z)-\phi^{t}_{S}(z))-8\sqrt{(1-r^{2})\eta}\bar{z}\bar{\eta}\phi^{t}_{S}(z)]\right.\\
&\left.\times\alpha_{s}(t_{c1})h_{c1}(\alpha_{c1}, \beta_{c1}, b_{3}, b_{B})\exp[-S_{B}(t_{c1})-S(t_{c1})-S_{3}(t_{c1})]\right.\\
&\left.+[\overline{\eta}(2-x_{B}-z-x_{3}\overline{\eta})\phi_{S}(z)+4\sqrt{(1-r^{2})\eta}\bar{z}\bar{\eta}(\overline{x}_{B}+z\overline{\eta}-2x_{3}\overline{\eta})(\phi^{s}_{S}(z)-\phi^{t}_{S}(z))\right.\\
&\left.+8\sqrt{(1-r^{2})\eta}r^{2}(\overline{x}_{B}-x_{3}\overline{\eta})\phi^{t}_{S}]\alpha_{s}(t_{d1})h_{d1}(\alpha_{d1}, \beta_{d1}, b_{3}, b_{B})\exp[-S_{B}(t_{d1})-S(t_{d1})-S_{3}(t_{d1})]\right\}
\end{split}
\end{eqnarray}

(2) $(V-A)(V+A)$

\begin{eqnarray}
\begin{split}
M^{LR}_{c}=&\frac{4\pi C_{F}M^{4}_{B}}{\sqrt{2N_{c}}}\int_{0}^{1}dx_{B}dzdx_{3}\int_{0}^{\infty}b_{B}db_{B}b_{3}db_{3}\phi_{B}(x_{B},b_{B})\\
&\left\{[4\sqrt{(1-r^{2})\eta}r(\overline{z}-x_{B}+\eta\overline{x}_{3})(\phi^{s}_{S}(z)-\phi^{t}_{S}(z))(\phi^{S}_{S}(x_{3})-\phi^{T}_{S}(x_{3}))+8\sqrt{(1-r^{2})\eta}r\overline{z}\right.\\
&\left.\times\phi^{T}_{S}(x_{3})(\phi^{s}_{S}(z)-\phi^{t}_{S}(z))-8\sqrt{(1-r^{2})\eta}r(x_{B}-\eta-x_{3}\overline{\eta})\phi^{t}_{S}(z)(\phi^{S}_{S}(x_{3})-\phi^{T}_{S}(x_{3}))\right.\\
&\left.-4r(-(x_{B}+\eta(z-2))+x_{3}\overline{\eta})\phi_{S}(z)(\phi^{S}_{S}(x_{3})-\phi^{T}_{S}(x_{3}))-8r\eta(\overline{z}+r^{2}(x_{3}-\overline{z}))\phi_{S}(z)\phi^{T}_{S}(x_{3})]\right.\\
&\left.\times\alpha_{s}(t_{c1})h_{c1}(\alpha_{c1}, \beta_{c1}, b_{3}, b_{B})\exp[-S_{B}(t_{c1})-S(t_{c1})-S_{3}(t_{c1})]\right.\\
&\left.+[4\sqrt{(1-r^{2})\eta}r(-2+x_{B}+z+x_{3}\overline{\eta})(\phi^{s}_{S}(z)-\phi^{t}_{S}(z))(\phi^{S}_{S}(x_{3})-\phi^{T}_{S}(x_{3}))+8\sqrt{(1-r^{2})\eta}r(\overline{x}_{B}-x_{3}\overline{\eta})\right.\\
&\left.\times\phi^{T}_{S}(x_{3})(\phi^{s}_{S}(z)-\phi^{t}_{S}(z))-4r(x_{3}\overline{\eta}-\overline{x}_{B}-\eta\overline{z})\phi_{S}(z)(\phi^{S}_{S}(x_{3})-\phi^{T}_{S}(x_{3}))+4r(r^{2}-1)(x_{3}\overline{\eta}-\overline{x}_{B})\phi_{S}(z)\phi^{T}_{S}(x_{3})]\right.\\
&\left.\times\alpha_{s}(t_{d1})h_{d1}(\alpha_{d1}, \beta_{d1}, b_{3}, b_{B})\exp[-S_{B}(t_{d1})-S(t_{d1})-S_{3}(t_{d1})]\right\}
\end{split}
\end{eqnarray}

(3) $(S-P)(S+P)$

\begin{eqnarray}
\begin{split}
M^{SP}_{c}=&\frac{-8\pi C_{F}M^{4}_{B}}{\sqrt{2N_{c}}}\int_{0}^{1}dx_{B}dzdx_{3}\int_{0}^{\infty}b_{B}db_{B}b_{3}db_{3}\phi_{B}(x_{B},b_{B})\phi_{S}(x_{3})\\
&\left\{[2\overline{\eta}(x_{3}-x_{B}+\overline{z}+\eta\overline{x}_{3})\phi_{S}(z)+2\sqrt{(1-r^{2})\eta}\bar{\eta}\bar{z} (\phi^{s}_{S}(z)-\phi^{t}_{S}(z))\right.\\
&\left.-4\sqrt{(1-r^{2})\eta}r^{2}(x_{B}-\eta-x_{3}\overline{\eta})\phi^{t}_{S}(z)]\right.\\
&\left.\times\alpha_{s}(t_{c1})h_{c1}(\alpha_{c1}, \beta_{c1}, b_{3}, b_{B})\exp[-S_{B}(t_{c1})-S(t_{c1})-S_{3}(t_{c1})]\right.\\
&\left.+[2\overline{\eta}(\overline{\eta}x_{3}-\eta\overline{z}-\overline{x}_{B})\phi_{S}(z)-2\sqrt{(1-r^{2})\eta}\bar{\eta}\bar{z}(\phi^{s}_{S}(z)+\phi^{t}_{S}(z))]\right.\\
&\left.\times\alpha_{s}(t_{d1})h_{d1}(\alpha_{d1}, \beta_{d1}, b_{3}, b_{B})\exp[-S_{B}(t_{d1})-S(t_{d1})-S_{3}(t_{d1})]\right\}
\end{split}
\end{eqnarray}

The contributions of the factorizable annihilation diagrams in Fig.~\ref{fig:figure1} (e1) and (f1) are

(1) $(V-A)(V-A)$

\begin{eqnarray}
\begin{split}
F^{LL}_{e}=&-2\pi C_{F}f_{B}M^{4}_{B}\int_{0}^{1}dzdx_{3}\int_{0}^{\infty}bdbb_{3}db_{3}\\
&\left\{[2\overline{\eta}(x_{3}\overline{\eta}-1)\phi_{S}(z)\phi_{S}(x_{3})-4\sqrt{(1-r^{2})\eta}r(-2+x_{3}\overline{\eta})\phi^{s}_{S}(z)\phi^{S}_{S}(x_{3})+4\sqrt{(1-r^{2})\eta}rx_{3}\overline{\eta}\phi^{s}_{S}(z)\phi^{T}_{S}(x_{3})]\right.\\
&\left.\times\alpha_{s}(t_{e1})h_{e1}(\alpha_{e1}, \beta_{e1}, b, b_{3})\exp[-S_{3}(t_{e1})-S(t_{e1})]S_{t}(x_{3})\right.\\
&\left.+[2z\overline{\eta}\phi_{S}(z)\phi_{S}(x_{3})-4\sqrt{(1-r^{2})\eta}r^{2}r(\overline{\eta}+z)\phi^{s}_{S}(z)\phi^{S}_{S}(x_{3})+4\sqrt{(1-r^{2})\eta}r^{2}r(\overline{\eta}-z)\phi^{t}_{S}(z)\phi^{S}_{S}(x_{3})]\right.\\
&\left.\times\alpha_{s}(t_{f1})h_{f1}(\alpha_{f1}, \beta_{f1}, b_{3}, b)\exp[-S_{3}(t_{f1})-S(t_{f1})]S_{t}(z)\right\}
\end{split}
\end{eqnarray}

(2) $(V-A)(V+A)$

\begin{eqnarray}
F^{LR}_{e}=F^{LL}_{e}
\end{eqnarray}

(3) $(S-P)(S+P)$

\begin{eqnarray}
\begin{split}
F^{SP}_{e}=&-4\pi C_{F}f_{B}M^{4}_{B}\int_{0}^{1}dzdx_{3}\int_{0}^{\infty}bdbb_{3}db_{3}\\
&\left\{[2r(x_{3}\overline{\eta}-1-\eta)\phi_{S}(z)(\phi^{S}_{S}(x_{3})-\phi^{T}_{S}(x_{3}))-4r(1-x_{3}\overline{\eta})\phi_{S}(z)\phi^{T}_{S}(x_{3})+4\sqrt{(1-r^{2})\eta}\overline{\eta}\phi^{s}_{S}(z)\phi_{S}(x_{3})]\right.\\
&\left.\times\alpha_{s}(t_{e1})h_{e1}(\alpha_{e1}, \beta_{e1}, b, b_{3})\exp[-S_{3}(t_{e1})-S(t_{e1})]S_{t}(x_{3})\right.\\
&\left.+[4r(\eta\overline{z}-1)\phi_{S}(z)\phi^{S}_{S}(x_{3})+2\sqrt{(1-r^{2})\eta}z\overline{\eta}\phi_{S}(x_{3})(\phi^{s}_{S}(z)-\phi^{t}_{S}(z))+4\sqrt{(1-r^{2})\eta}r^{2}\overline{\eta}\phi_{S}(x_{3})\phi^{t}_{S}(z)]\right.\\
&\left.\times\alpha_{s}(t_{f1})h_{f1}(\alpha_{f1}, \beta_{f1}, b_{3}, b)\exp[-S_{3}(t_{f1})-S(t_{f1})]S_{t}(z)\right\}
\end{split}
\end{eqnarray}

The contributions of the nonfactorizable annihilation diagrams in Fig.~\ref{fig:figure1} (g1) and (h1) are

(1) $(V-A)(V-A)$

\begin{eqnarray}
\begin{split}
M^{LL}_{g}=&\frac{4\pi C_{F}M^{4}_{B}}{\sqrt{2N_{c}}}\int_{0}^{1}dx_{B}dzdx_{3}\int_{0}^{\infty}b_{B}db_{B}b_{3}db_{3}\phi_{B}(x_{B},b_{B})\\
&\left\{[4(1+\eta)z\overline{\eta}\phi_{S}(z)\phi_{S}(x_{3})+4\sqrt{(1-r^{2})\eta}r(x_{B}-z-\bar{x}_{3}\bar{\eta})(\phi^{s}_{S}(z)\phi^{S}_{S}(x_{3})-\phi^{t}_{S}(z)\phi^{T}_{S}(x_{3}))+4\sqrt{(1-r^{2})\eta}r\right.\\
&\left.\times(x_{B}+z-\bar{x}_{3}\bar{\eta})(\phi_{S}(z)\phi^{T}_{S}(x_{3})-\phi^{t}_{S}(z)\phi^{S}_{S}(x_{3}))]\alpha_{s}(t_{g1})h_{g1}(\alpha_{g1}, \beta_{g1}, b_{B}, b_{3})\exp[-S_{B}(t_{g1})-S(t_{g1})-S_{3}(t_{g1})]\right.\\
&\left.+[4\overline{\eta}(x_{3}\overline{\eta}-(x_{B}+(z-2)\eta)+r_{b})\phi_{S}(z)\phi_{S}(x_{3})-4\sqrt{(1-r^{2})\eta}r(\overline{z}+x_{3}\overline{\eta}+\eta-x_{B})(\phi^{s}_{S}(z)\phi^{S}_{S}(x_{3})-\phi^{t}_{S}(z)\phi^{T}_{S}(x_{3}))\right.\\
&\left.-4\sqrt{(1-r^{2})\eta}r(\overline{z}-x_{3}\overline{\eta}+x_{B}-\eta)(\phi_{S}(z)\phi^{T}_{S}(x_{3})-\phi^{t}_{S}(z)\phi^{S}_{S}(x_{3}))+16\sqrt{(1-r^{2})\eta}rr_{b}\phi^{s}_{S}(z)\phi^{S}_{S}(x_{3})]\right.\\
&\left.\times\alpha_{s}(t_{g1})h_{g1}(\alpha_{h1}, \beta_{h1}, b_{B}, b_{3})\exp[-S_{B}(t_{h1})-S(t_{h1})-S_{3}(t_{h1})]\right\}
\end{split}
\end{eqnarray}

(2) $(V-A)(V+A)$

\begin{eqnarray}
\begin{split}
M^{LR}_{g}=&\frac{-4\pi C_{F}M^{4}_{B}}{\sqrt{2N_{c}}}\int_{0}^{1}dx_{B}dzdx_{3}\int_{0}^{\infty}b_{B}db_{B}b_{3}db_{3}\phi_{B}(x_{B},b_{B})\\
&\left\{[-4r(x_{3}\overline{\eta}+\eta\overline{z}-\overline{x}_{B})\phi_{S}(z)(\phi^{S}_{S}(x_{3})-\phi^{T}_{S}(x_{3}))+8rz\eta\phi_{S}(z)\phi^{T}_{S}(x_{3})+4\sqrt{(1-r^{2})\eta}(z\overline{\eta}+\eta\overline{z}+\eta)\right.\\
&\left.\times\phi_{S}(x_{3})(\phi^{s}_{S}(z)-\phi^{t}_{S}(z))+8\sqrt{(1-r^{2})\eta}z\overline{\eta}\phi_{S}(x_{3})\phi^{t}_{S}(z)]\alpha_{s}(t_{g1})h_{g1}(\alpha_{g1}, \beta_{g1}, b_{B}, b_{3})\right.\\
&\left.\times\exp[-S_{B}(t_{g1})-S(t_{g1})-S_{3}(t_{g1})]+[4r(x_{3}\overline{\eta}-x_{B}+\eta(2-z))\phi_{S}(z)(\phi^{S}_{S}(x_{3})-\phi^{T}_{S}(x_{3}))+8r\eta\overline{z}\phi_{S}(z)\phi^{T}_{S}(x_{3})\right.\\
&\left.+4\sqrt{(1-r^{2})\eta}\overline{\eta}(\overline{z}+1)\phi_{S}(x_{3})(\phi^{s}_{S}(z)+\phi^{t}_{S}(z))+4rr_{b}(1+\eta)\phi_{S}(z)(\phi^{S}_{S}(x_{3})-\phi^{T}_{S}(x_{3}))+8rr_{b}\eta\phi_{S}(z)\phi^{T}_{S}(x_{3})]\right.\\
&\left.\times\alpha_{s}(t_{g1})h_{g1}(\alpha_{h1}, \beta_{h1}, b_{B}, b_{3})\exp[-S_{B}(t_{h1})-S(t_{h1})-S_{3}(t_{h1})]\right\}
\end{split}
\end{eqnarray}

(3) $(S-P)(S+P)$

\begin{eqnarray}
\begin{split}
M^{SP}_{g}=&\frac{8\pi C_{F}M^{4}_{B}}{\sqrt{2N_{c}}}\int_{0}^{1}dx_{B}dzdx_{3}\int_{0}^{\infty}b_{B}db_{B}b_{3}db_{3}\phi_{B}(x_{B},b_{B})\\
&\left\{[2\overline{\eta}(\eta\overline{z}+x_{3}\overline{\eta}-\overline{x}_{B})\phi_{S}(z)\phi_{S}(x_{3})+2\sqrt{(1-r^{2})\eta}r(x_{B}-z-\overline{x}_{3}\overline{\eta})(\phi^{t}_{S}(z)\phi^{T}_{S}(x_{3})-\phi^{s}_{S}(z)\phi^{S}_{S}(x_{3}))\right.\\
&\left.+2\sqrt{(1-r^{2})\eta}r(x_{B}+z-\overline{x}_{3}\overline{\eta})(\phi^{s}_{S}(z)\phi^{T}_{S}(x_{3})-\phi^{t}_{S}(z)\phi^{S}_{S}(x_{3}))]\alpha_{s}(t_{g1})h_{g1}(\alpha_{g1}, \beta_{g1}, b_{B}, b_{3})\right.\\
&\left.\times\exp[-S_{B}(t_{g1})-S(t_{g1})-S_{3}(t_{g1})]+[2(-\bar{z}\bar{\eta}(1+\eta)+r_{b}\overline{\eta})\phi_{S}(z)\phi_{S}(x_{3})-2\sqrt{(1-r^{2})\eta}r(\eta-x_{B}+x_{3}\overline{\eta}+\overline{z})\right.\\
&\left.\times(\phi^{t}_{S}(z)\phi^{T}_{S}(x_{3})-\phi^{s}_{S}(z)\phi^{S}_{S}(x_{3}))-2\sqrt{(1-r^{2})\eta}r(x_{B}-\eta-x_{3}\overline{\eta}+\overline{z})(\phi^{s}_{S}(z)\phi^{T}_{S}(x_{3})-\phi^{t}_{S}(z)\phi^{S}_{S}(x_{3}))]\right.\\
&\left.\times\alpha_{s}(t_{h1})h_{h1}(\alpha_{h1}, \beta_{h1}, b_{B}, b_{3})\exp[-S_{B}(t_{h1})-S(t_{h1})-S_{3}(t_{h1})]\right\}
\end{split}
\end{eqnarray}

In Fig.~\ref{fig:figure1} (a2) and (b2), we can see the meson pair will be factorized out, when taking the $(V-A)(V-A)$ and $(V-A)(V+A)$ currents into account, the $S$-wave meson pair cannot be emitted because of the charge conjugation invariance, so

\begin{eqnarray}
F^{'LL}_{a}=F^{'LR}_{a}=0
\end{eqnarray}

\begin{eqnarray}
\begin{split}
F^{'SP}_{a}=&8\pi C_{F}\overline{f}_{s}\sqrt{\eta}M^{4}_{B}r\int_{0}^{1}dx_{B}dx_{3}\int_{0}^{\infty}b_{B}db_{B}b_{3}db_{3}\phi_{B}(x_{B},b_{B})\\
&\left\{[2\overline{\eta}(r_{b}-2)\phi_{S}(x_{3})-2r(1+\eta+x_{3}\overline{\eta})(\phi^{S}_{S}(x_{3})-\phi^{T}_{S}(x_{3}))-4r\phi^{T}_{S}(x_{3})+8rr_{b}\phi^{S}_{S}(x_{3})]\right.\\
&\left.\times\alpha_{s}(t_{a2})h_{a2}(\alpha_{a2}, \beta_{a2}, b_{B}, b_{3})\exp[-S_{B}(t_{a2})-S_{3}(t_{a2})]S_{t}(x_{3})\right.\\
&\left.+[4r(\overline{\eta}-x_{B})\phi^{S}_{S}(x_{3})+2\overline{\eta}\phi_{S}(x_{3})]\alpha_{s}(t_{b2})h_{b2}(\alpha_{b2}, \beta_{b2}, b_{3}, b_{B})\exp[-S_{B}(t_{b2})-S_{3}(t_{b2})]S_{t}(x_{B})\right\}
\end{split}
\end{eqnarray}

The nonfactorizable diagrams in Fig.~\ref{fig:figure1} (c2) and (d2) yield

(1)$(V-A)(V-A)$

\begin{eqnarray}
\begin{split}
M^{'LL}_{c}=&\frac{4\pi C_{F}M^{4}_{B}}{\sqrt{2N_{c}}}\int_{0}^{1}dx_{B}dzdx_{3}\int_{0}^{\infty}b_{B}db_{B}bdb\phi_{B}(x_{B},b_{B})\phi_{S}(z)\\
&\left\{[4\overline{\eta}^{2}(x_{B}-z)\phi_{S}(x_{3})+4r(x_{3}\overline{\eta}+\eta\overline{z}-\overline{x}_{B})(\phi^{S}_{S}(x_{3})-\phi^{T}_{S}(x_{3}))+8r\eta(x_{B}-z)\phi^{T}_{S}(x_{3})]\right.\\
&\left.\times\alpha_{s}(t_{c2})h_{c2}(\alpha_{c2}, \beta_{c2}, b, b_{B})\exp[-S_{B}(t_{c2})-S(t_{c2})-S_{3}(t_{c2})]\right.\\
&\left.+[4\overline{\eta}(\overline{z}+\overline{x}_{B}-x_{3}\overline{\eta})\phi_{S}(x_{3})+4r(x_{3}\overline{\eta}-\eta\overline{z}-\overline{x}_{B})(\phi^{S}_{S}(x_{3})-\phi^{T}_{S}(x_{3}))+8r(1-x_{3}\overline{\eta})\phi^{T}_{S}(x_{3})]\right.\\
&\left.\times\alpha_{s}(t_{d2})h_{d2}(\alpha_{d2}, \beta_{d2}, b, b_{B})\exp[-S_{B}(t_{d2})-S(t_{d2})-S_{3}(t_{d2})]\right\}
\end{split}
\end{eqnarray}

(2)$(V-A)(V+A)$

\begin{eqnarray}
\begin{split}
M^{'LR}_{c}=&\frac{4\pi C_{F}M^{4}_{B}}{\sqrt{2N_{c}}}\int_{0}^{1}dx_{B}dzdx_{3}\int_{0}^{\infty}b_{B}db_{B}bdb\phi_{B}(x_{B},b_{B})\\
&\left\{[4\sqrt{(1-r^{2})\eta}(\overline{x}_{3}\overline{\eta}+z-x_{B})(\phi^{S}_{S}(x_{3})-\phi^{T}_{S}(x_{3}))(\phi^{s}_{S}(z)-\phi^{t}_{S}(z))-8\sqrt{(1-r^{2})\eta}r(x_{B}-z)\right.\\
&\left.\times\phi^{t}_{S}(z)(\phi^{S}_{S}(x_{3})-\phi^{T}_{S}(x_{3}))+8\sqrt{(1-r^{2})\eta}r\overline{x}_{3}\overline{\eta}\phi^{T}_{S}(x_{3})(\phi^{s}_{S}(z)-\phi^{t}_{S}(z))-4\sqrt{(1-r^{2})\eta}\overline{\eta}(x_{B}-z)\right.\\
&\left.\times\phi_{S}(x_{3})(\phi^{s}_{S}(z)-\phi^{t}_{S}(z))+8\sqrt{(1-r^{2})\eta}\overline{\eta}(x_{B}-z)\phi_{S}(x_{3})\phi^{t}_{S}(z)]\alpha_{s}(t_{c2})h_{c2}(\alpha_{c2}, \beta_{c2}, b, b_{B})\right.\\
&\left.\times\exp[-S_{B}(t_{c2})-S(t_{c2})-S_{3}(t_{c2})]+[4\sqrt{(1-r^{2})\eta}r(x_{3}\overline{\eta}-\overline{x}_{B}-\overline{z})(\phi^{S}_{S}(x_{3})-\phi^{T}_{S}(x_{3}))(\phi^{s}_{S}(z)-\phi^{t}_{S}(z))\right.\\
&\left.-8\sqrt{(1-r^{2})\eta}r(1-x_{3}\overline{\eta})\phi^{t}_{S}(z)(\phi^{S}_{S}(x_{3})-\phi^{T}_{S}(x_{3}))-8\sqrt{(1-r^{2})\eta}r(x_{B}-\overline{z})\phi^{T}_{S}(x_{3})(\phi^{s}_{S}(z)-\phi^{t}_{S}(z))\right.\\
&\left.+4\sqrt{(1-r^{2})\eta}\overline{\eta}(\overline{z}-x_{B})\phi_{S}(x_{3})(\phi^{s}_{S}(z)-\phi^{t}_{S}(z))+8\sqrt{(1-r^{2})\eta}(1-x_{3}\overline{\eta})\phi_{S}(x_{3})\phi^{t}_{S}(z)]\right.\\
&\left.\times\alpha_{s}(t_{d2})h_{d2}(\alpha_{d2}, \beta_{d2}, b, b_{B})\exp[-S_{B}(t_{d2})-S(t_{d2})-S_{3}(t_{d2})]\right\}
\end{split}
\end{eqnarray}

(3) $(S-P)(S+P)$

\begin{eqnarray}
\begin{split}
M^{'SP}_{c}=&\frac{-8\pi C_{F}M^{4}_{B}}{\sqrt{2N_{c}}}\int_{0}^{1}dx_{B}dzdx_{3}\int_{0}^{\infty}b_{B}db_{B}bdb\phi_{B}(x_{B},b_{B})\phi_{S}(z)\\
&\left\{[2\overline{\eta}(\overline{x}_{3}\overline{\eta}+z-x_{B})\phi_{S}(x_{3})-2r(\eta(x_{B}-z)-\overline{x}_{3}\overline{\eta})(\phi^{S}_{S}(x_{3})-\phi^{T}_{S}(x_{3}))+4r\overline{x}_{3}\overline{\eta}\phi^{T}_{S}(x_{3})]\right.\\
&\left.\times\alpha_{s}(t_{c2})h_{c2}(\alpha_{c2}, \beta_{c2}, b, b_{B})\exp[-S_{B}(t_{c2})-S(t_{c2})-S_{3}(t_{c2})]\right.\\
&\left.+[2\overline{\eta}(1+\eta)(x_{B}-z)\phi_{S}(x_{3})+2r(x_{3}\overline{\eta}-\eta(\overline{x}_{B}+\overline{z})-1)(\phi^{S}_{S}(x_{3})-\phi^{T}_{S}(x_{3}))+4r\eta(z-\overline{x}_{B})\phi^{T}_{S}(x_{3})]\right.\\
&\left.\times\alpha_{s}(t_{d2})h_{d2}(\alpha_{d2}, \beta_{d2}, b, b_{B})\exp[-S_{B}(t_{d2})-S(t_{d2})-S_{3}(t_{d2})]\right\}
\end{split}
\end{eqnarray}

The contributions of the factorizable annihilation diagrams in Fig.~\ref{fig:figure1} (e2) and (f2) are

(1)$(V-A)(V-A)$

\begin{eqnarray}
\begin{split}
F^{'LL}_{e}=&2\pi C_{F}f_{B}M^{4}_{B}\int_{0}^{1}dzdx_{3}\int_{0}^{\infty}bdbb_{3}db_{3}\\
&\left\{[-2\bar{z}\bar{\eta}\phi_{S}(x_{3})\phi_{S}(z)+4\sqrt{(1-r^{2})\eta}r(2-z)\phi^{S}_{S}(x_{3})\phi^{s}_{S}(z)+4\sqrt{(1-r^{2})\eta}rz\phi^{S}_{S}(x_{3})\phi^{t}_{S}(z)]\right.\\
&\left.\times\alpha_{s}(t_{e2})h_{e2}(\alpha_{e2}, \beta_{e2}, b_{3}, b)\exp[-S_{3}(t_{e2})-S(t_{e2})]S_{t}(z)\right.\\
&\left.+[2\overline{\eta}(\eta+x_{3}\overline{\eta})\phi_{S}(x_{3})\phi_{S}(z)-4\sqrt{(1-r^{2})\eta}r(1+x_{3}\overline{\eta}+\eta)\phi^{S}_{S}(x_{3})\phi^{s}_{S}(z)+4\sqrt{(1-r^{2})\eta}r\overline{x}_{3}\overline{\eta}\right.\\
&\left.\times\phi^{T}_{S}(x_{3})\phi^{s}_{S}(z)]\alpha_{s}(t_{f2})h_{f2}(z, x_{3}, b, b_{3})\exp[-S_{3}(t_{f2})-S(t_{f2})]S_{t}(x_{3})\right\}
\end{split}
\end{eqnarray}

(2) $(V-A)(V+A)$

\begin{eqnarray}
F^{'LR}_{e}=F^{'LL}_{e}
\end{eqnarray}

\begin{eqnarray}
\begin{split}
F^{'SP}_{e}=&-4\pi C_{F}f_{B}M^{4}_{B}\int_{0}^{1}dzdx_{3}\int_{0}^{\infty}bdbb_{3}db_{3}\\
&\left\{[-2\sqrt{(1-r^{2})\eta}\overline{z}\phi_{S}(x_{3})(\phi^{s}_{S}(z)-\phi^{t}_{S}(z))-4\sqrt{(1-r^{2})\eta}\bar{z}\bar{\eta}\phi_{S}(x_{3})\phi^{t}_{S}(z)+4r(1+\eta\overline{z})\right.\\
&\left.\times\phi^{S}_{S}(x_{3})\phi_{S}(z)]\alpha_{s}(t_{e2})h_{e2}(\alpha_{e2}, \beta_{e2}, b_{3}, b)\exp[-S_{3}(t_{e2})-S(t_{e2})]S_{t}(z)\right.\\
&\left.+[-4\sqrt{(1-r^{2})\eta}\overline{\eta}\phi_{S}(x_{3})\phi^{s}_{S}(z)+2r(x_{3}\overline{\eta}+2\eta)\phi_{S}(z)(\phi^{S}_{S}(x_{3})-\phi^{T}_{S}(x_{3}))+4r\eta\phi_{S}(z)\phi^{T}_{S}(x_{3})]\right.\\
&\left.\times\alpha_{s}(t_{f2})h_{f2}(\alpha_{f2}, \beta_{f2}, b, b_{3})\exp[-S_{3}(t_{f2})-S(t_{f2})]S_{t}(x_{3})\right\}
\end{split}
\end{eqnarray}

The contributions of the nonfactorizable annihilation diagrams in Fig.~\ref{fig:figure1} (g2) and (h2) are

(1) $(V-A)(V-A)$

\begin{eqnarray}
\begin{split}
M^{'LL}_{g}=&\frac{-4\pi C_{F}M^{4}_{B}}{\sqrt{2N_{c}}}\int_{0}^{1}dx_{B}dzdx_{3}\int_{0}^{\infty}b_{B}db_{B}bdb\phi_{B}(x_{B},b_{B})\\
&\left\{[4\overline{\eta}((2-z)\eta+x_{3}\overline{\eta}-x_{B})\phi_{S}(z)\phi_{S}(x_{3})+4\sqrt{(1-r^{2})\eta}r(\overline{z}+x_{3}\overline{\eta}+\eta-x_{B})(\phi^{T}_{S}(x_{3})\phi^{t}_{S}(z)\right.\\
&\left.-\phi^{S}_{S}(x_{3})\phi^{s}_{S}(z))+4\sqrt{(1-r^{2})\eta}r(\overline{z}-x_{3}\overline{\eta}-\eta+x_{B})(\phi^{T}_{S}(x_{3})\phi^{s}_{S}(z)-\phi^{S}_{S}(x_{3})\phi^{t}_{S}(z))]\right.\\
&\left.\times\alpha_{s}(t_{g2})h_{g2}(\alpha_{g2}, \beta_{g2}, b_{B}, b)\exp[-S_{B}(t_{g2})-S(t_{g2})-S_{3}(t_{g2})]\right.\\
&\left.+[4(z\overline{\eta}(1+\eta)+r_{b}\overline{\eta})\phi_{S}(z)\phi_{S}(x_{3})-4\sqrt{(1-r^{2})\eta}r(x_{B}-z-\overline{x}_{3}\overline{\eta})(\phi^{T}_{S}(x_{3})\phi^{t}_{S}(z)\right.\\
&\left.-\phi^{S}_{S}(x_{3})\phi^{s}_{S}(z))-4\sqrt{(1-r^{2})\eta}r(x_{B}+z-\overline{x}_{3}\overline{\eta})(\phi^{T}_{S}(x_{3})\phi^{s}_{S}(z)-\phi^{S}_{S}(x_{3})\phi^{t}_{S}(z))+16\sqrt{(1-r^{2})\eta}r_{b}r\right.\\
&\left.\times\phi^{S}_{S}(x_{3})\phi^{s}_{S}(z)]\alpha_{s}(t_{h2})h_{h2}(\alpha_{h2}, \beta_{h2}, b_{B}, b)\exp[-S_{B}(t_{h2})-S(t_{h2})-S_{3}(t_{h2})]\right\}
\end{split}
\end{eqnarray}

(2) $(V-A)(V+A)$

\begin{eqnarray}
\begin{split}
M^{'LR}_{g}=&\frac{-4\pi C_{F}M^{4}_{B}}{\sqrt{2N_{c}}}\int_{0}^{1}dx_{B}dzdx_{3}\int_{0}^{\infty}b_{B}db_{B}bdb\phi_{B}(x_{B},b_{B})\\
&\left\{[4\sqrt{(1-r^{2})\eta}\bar{z}\bar{\eta}\phi_{S}(x_{3})(\phi^{s}_{S}(z)-\phi^{t}_{S}(z))-8\sqrt{(1-r^{2})\eta}r^{2}(x_{B}-\eta-x_{3}\overline{\eta})\phi_{S}(x_{3})\phi^{t}_{S}(z)+4r(\eta(2-z)-x_{B}\right.\\
&\left.+x_{3}\overline{\eta})\phi_{S}(z)(\phi^{S}_{S}(x_{3})-\phi^{T}_{S}(x_{3}))-8r(x_{B}-\eta-x_{3}\overline{\eta})\phi^{T}_{S}(x_{3})\phi_{S}(z)]\alpha_{s}(t_{g2})h_{g2}(\alpha_{g2}, \beta_{g2}, b_{B}, b)\right.\\
&\left.\times\exp[-S_{B}(t_{g2})-S(t_{g2})-S_{3}(t_{g2})]+[4\sqrt{(1-r^{2})\eta}\overline{\eta}(z+r_{b})\phi_{S}(x_{3})(\phi^{s}_{S}(z)-\phi^{t}_{S}(z))-4\sqrt{(1-r^{2})\eta}(x_{3}\overline{\eta}\right.\\
&\left.+\eta\overline{z}-\overline{x}_{B})\phi_{S}(z)(\phi^{S}_{S}(x_{3})-\phi^{T}_{S}(x_{3}))-8\sqrt{(1-r^{2})\eta}r^{2}(x_{B}-\overline{x}_{3}\overline{\eta}-r_{b})\phi_{S}(x_{3})\phi^{t}_{S}(z)+8r\right.\\
&\left.\times(x_{B}-\overline{x}_{3}\overline{\eta})\phi_{S}(x_{3})\phi^{t}_{S}(z)+4r(1+\eta)\phi_{S}(z)(\phi^{S}_{S}(x_{3})+\phi^{T}_{S}(x_{3}))]\right.\\
&\left.\times\alpha_{s}(t_{h2})h_{h2}(\alpha_{h2}, \beta_{h2}, b_{B}, b)\exp[-S_{B}(t_{h2})-S(t_{h2})-S_{3}(t_{h2})]\right\}
\end{split}
\end{eqnarray}

(3) $(S-P)(S+P)$

\begin{eqnarray}
\begin{split}
M^{'SP}_{g}=&\frac{8\pi C_{F}M^{4}_{B}}{\sqrt{2N_{c}}}\int_{0}^{1}dx_{B}dzdx_{3}\int_{0}^{\infty}b_{B}db_{B}bdb\phi_{B}(x_{B},b_{B})\\
&\left\{[-2\bar{z}\bar{\eta}(1+\eta)\phi_{S}(x_{3})\phi_{S}(z)+2\sqrt{(1-r^{2})\eta}r(x_{3}\overline{\eta}+\overline{z}-x_{B}+\eta)(\phi^{S}_{S}(x_{3})\phi^{s}_{S}(z)-\phi^{T}_{S}(x_{3})\phi^{t}_{S}(z))\right.\\
&\left.+2\sqrt{(1-r^{2})\eta}r(-x_{3}\overline{\eta}+\overline{z}+x_{B}-\eta)(\phi^{T}_{S}(x_{3})\phi^{s}_{S}(z)-\phi^{S}_{S}(x_{3})\phi^{t}_{S}(z))]\alpha_{s}(t_{g2})h_{g2}(\alpha_{g2}, \beta_{g2}, b_{B}, b)\right.\\
&\left.\times\exp[-S_{B}(t_{g2})-S(t_{g2})-S_{3}(t_{g2})]+[2\overline{\eta}(x_{3}\overline{\eta}+\eta\overline{z}-\overline{x}_{B}+r_{b})\phi_{S}(z)\phi_{S}(x_{3})-2\sqrt{(1-r^{2})\eta}r\right.\\
&\left.\times(x_{B}-z-\overline{x}_{3}\overline{z})(\phi^{S}_{S}(x_{3})\phi^{s}_{S}(z)-\phi^{T}_{S}(x_{3})\phi^{t}_{S}(z))-2\sqrt{(1-r^{2})\eta}r(x_{B}+z-\overline{x}_{3}\overline{z})(\phi^{T}_{S}(x_{3})\phi^{s}_{S}(z)-\phi^{S}_{S}(x_{3})\phi^{t}_{S}(z))]\right.\\
&\left.\times\alpha_{s}(t_{h2})h_{h2}(\alpha_{h2}, \beta_{h2}, b_{B}, b)\exp[-S_{B}(t_{h2})-S(t_{h2})-S_{3}(t_{h2})]\right\}
\end{split}
\end{eqnarray}

The hard functions $h_{i}$ are derived from the Fourier transform with $i=(a1, ... ,h2)$, whose specific expression are

\begin{eqnarray}
\begin{split}
h_{i}(\alpha, \beta, b_{1}, b_{2})&=h_{1}(\alpha, b_{1})\times h_{2}(\beta, b_{1}, b_{2}),\\
h_{1}(\alpha, b_{1})&=
\begin{cases}
K_{0}(\sqrt{\alpha}b_{1}),&\alpha\textgreater 0,\\
K_{0}(i\sqrt{-\alpha}b_{1}),&\alpha\textless 0,
\end{cases}\\
h_{2}(\beta, b_{1}, b_{2})&=
\begin{cases}
\theta(b_{1}-b_{2})I_{0}(\sqrt{\beta}b_{2})K_{0}(\sqrt{\beta}b_{1})+(b_{1}\leftrightarrow b_{2}),&\beta\textgreater 0,\\
\theta(b_{1}-b_{2})J_{0}(\sqrt{-\beta}b_{2})K_{0}(i\sqrt{-\beta}b_{1})+(b_{1}\leftrightarrow b_{2}),&\beta\textless 0,
\end{cases}
\end{split}
\end{eqnarray}
with the Bessel function $J_{0}$, the modified Bessel functions $K_{0}$ and $I_{0}$. The expressions for $\alpha$ and $\beta$ in the hard functions are

\begin{eqnarray}
\begin{split}
\alpha_{(a1, b1)}=\beta_{(c1, d1)}&=M^{2}_{B}\overline{z}(1-r^{2})(x_{B}-\eta),\\
\beta_{a1}&=M^{2}_{B}(1-\overline{\eta}(z+r^{2}\overline{z})),\\
\beta_{b1}&=M^{2}_{B}(1-r^{2})(x_{B}-\eta),\\
\alpha_{c1}&=M^{2}_{B}(\overline{z}(1-r^{2})+r^{2}x_{3})(x_{B}-\eta-x_{3}\overline{\eta}),\\
\alpha_{d1}&=M^{2}_{B}(r^{2}(z-x_{3})+\overline{z})(x_{3}\overline{\eta}-\overline{x}_{B}),\\
\alpha_{(e1, f1)}=\beta_{(g1, h1)}&=M^{2}_{B}\overline{x}_{3}\overline{\eta}(z(r^{2}-1)-r^{2}\overline{x}_{3}),\\
\beta_{e1}&=M^{2}_{B}(r^{2}x_{3}-1)(1-x_{3}\overline{\eta}),\\
\beta_{f1}&=M^{2}_{B}\overline{\eta}(-r^{2}\overline{z}-z),\\
\alpha_{g1}&=M^{2}_{B}(z(1-r^{2})+r^{2}\overline{x}_{3})(x_{B}-\overline{x}_{3}\overline{\eta}),\\
\alpha_{h1}&=M^{2}_{B}(1+(\overline{z}(1-r^{2})+r^{2}x_{3})(x_{B}-\eta-x_{3}\overline{\eta})),\\
\alpha_{(a2, b2)}=\beta_{(c2, d2)}&=M^{2}_{B}\overline{x}_{3}\overline{\eta}(x_{B}-r^{2}\overline{x}_{3}),\\
\beta_{a2}&=M^{2}_{B}(1-(1-r^{2}\overline{x}_{3})(x_{3}\overline{\eta}+\eta)),\\
\beta_{b2}&=M^{2}_{B}\overline{\eta}(x_{B}-r^{2}),\\
\alpha_{c2}&=M^{2}_{B}\overline{x}_{3}\overline{\eta}(x_{B}-z+r^{2}(z-\overline{x}_{3})),\\
\alpha_{d2}&=M^{2}_{B}(1-x_{3}\overline{\eta})(r^{2}x_{3}+z(1-r^{2})-\overline{x}_{B}),\\
\alpha_{(e2, f2)}=\beta_{(g2, h2)}&=M^{2}_{B}(x_{3}\overline{\eta}+\eta)(-\overline{z}(1-r^{2})-r^{2}x_{3}),\\
\beta_{e2}&=M^{2}_{B}(z(1-r^{2})-1),\\
\beta_{f2}&=M^{2}_{B}(1-r^{2}\overline{x}_{3})(-\eta-x_{3}\overline{\eta}),\\
\alpha_{g2}&=M^{2}_{B}(\overline{z}(1-r^{2})+r^{2}x_{3})(x_{B}-\eta-x_{3}\overline{\eta}),\\
\alpha_{h2}&=M^{2}_{B}(1+(z(1-r^{2})+r^{2}\overline{x}_{3})(x_{B}-\overline{x}_{3}\overline{\eta}))
\end{split}
\end{eqnarray}

The hard scales $t_{i}(i=a1,...h2)$, which are taken to remove the large logarithmic radiative corrections, are given by

\begin{eqnarray}
\begin{split}
t_{a1}&=\mathrm{Max}\left\{\sqrt{\left|\beta_{a1}\right|},1/b_{B}, 1/b\right\},t_{b1}=\mathrm{Max}\left\{\sqrt{\left|\beta_{b1}\right|},1/b_{B}, 1/b\right\},\\
t_{c1}&=\mathrm{Max}\left\{\sqrt{\left|\alpha_{c1}\right|}, \sqrt{\left|\beta_{c1}\right|},1/b_{B}, 1/b_{3}\right\},t_{d1}=\mathrm{Max}\left\{\sqrt{\left|\alpha_{d1}\right|}, \sqrt{\left|\beta_{d1}\right|},1/b_{B}, 1/b_{3}\right\},\\
t_{e1}&=\mathrm{Max}\left\{\sqrt{\left|\beta_{e1}\right|},1/b, 1/b_{3}\right\},t_{f1}=\mathrm{Max}\left\{\sqrt{\left|\beta_{f1}\right|},1/b, 1/b_{3}\right\},\\
t_{g1}&=\mathrm{Max}\left\{\sqrt{\left|\alpha_{g1}\right|}, \sqrt{\left|\beta_{g1}\right|},1/b_{B}, 1/b_{3}\right\},t_{h1}=\mathrm{Max}\left\{\sqrt{\left|\alpha_{h1}\right|}, \sqrt{\left|\beta_{h1}\right|},1/b_{B}, 1/b_{3}\right\},\\
t_{a2}&=\mathrm{Max}\left\{\sqrt{\left|\beta_{a2}\right|},1/b_{B}, 1/b_{3}\right\},t_{b2}=\mathrm{Max}\left\{\sqrt{\left|\beta_{b2}\right|},1/b_{B}, 1/b_{3}\right\},\\
t_{c2}&=\mathrm{Max}\left\{\sqrt{\left|\alpha_{c2}\right|}, \sqrt{\left|\beta_{c2}\right|},1/b_{B}, 1/b\right\},t_{d2}=\mathrm{Max}\left\{\sqrt{\left|\alpha_{d2}\right|}, \sqrt{\left|\beta_{d2}\right|},1/b_{B}, 1/b\right\},\\
t_{e2}&=\mathrm{Max}\left\{\sqrt{\left|\beta_{e2}\right|},1/b, 1/b_{3}\right\},t_{f2}=\mathrm{Max}\left\{\sqrt{\left|\beta_{f2}\right|},1/b, 1/b_{3}\right\},\\
t_{g2}&=\mathrm{Max}\left\{\sqrt{\left|\alpha_{g2}\right|}, \sqrt{\left|\beta_{g2}\right|},1/b_{B}, 1/b\right\},t_{h2}=\mathrm{Max}\left\{\sqrt{\left|\alpha_{h2}\right|}, \sqrt{\left|\beta_{h2}\right|},1/b_{B}, 1/b\right\}.\\
\end{split}
\end{eqnarray}

The Sudakov exponents are defined as

\begin{eqnarray}
\begin{split}
&S_{B}=s(x_{B}p^+_{1},b_{B})+\frac{5}{3}\int^{t}_{1/b_{B}}\frac{\emph{d}{\bar{\mu}}}{\bar{\mu}}\gamma_{q}(\alpha_{s}({\bar{\mu}})),\\
&S=s(zp^{+},b)+s(\bar{z}p^+,b)+2\int^{t}_{1/b}\frac{\emph{d}{\bar{\mu}}}{\bar{\mu}}\gamma_{q}(\alpha_{s}({\bar{\mu}})),\\
&S_{3}=s({x}_{3}p^{-}_{3},b_{3})+s(\bar{x}_{3}p^-_{3},b_{3})+2\int^{t}_{1/b_{3}}\frac{\emph{d}{\bar{\mu}}}{\bar{\mu}}\gamma_{q}(\alpha_{s}({\bar{\mu}})),
\end{split}
\end{eqnarray}
with the anomalous dimension of the quark $\gamma_{q}=-\alpha_{s}/\pi$, and $s(Q,b)$ is the Sudakov factor, which can be found in Ref.~\cite{H:2003fha}. Meanwhile, the threshold resummation factor $S_{t}(x)$ is taken from Ref.~\cite{Kurimoto:2001xti},

\begin{equation}
S_{t}(x)=\frac{2^{1+2c}\Gamma(\frac{3}{2}+c)}{\sqrt{\pi}\Gamma(1+c)}[x(1-x)]^{c},
\end{equation}
with the parameter $c=0.3$.

%%%--=================================================================
%%%=====           Reference        ==========================
%%5===================================================================

\end{document}